\documentclass{jfm}
\usepackage{graphicx}
\usepackage{epstopdf, epsfig}
\usepackage{tikz}
\usepackage{amsmath}
\usepackage[utf8]{inputenc}
\usepackage{booktabs}
\usepackage{lscape}

\usepackage{rotating}

\usepackage{xcolor}

\newcommand{\tikzcircle}[2][white,fill=red]{\tikz[baseline=-0.5ex]\draw[#1,radius=#2] (0,0.01) circle ;}%
\definecolor{c6}{RGB}{0,0,0}
\definecolor{c5}{RGB}{230,159,0}
\definecolor{c4}{RGB}{196, 0, 96}
\definecolor{c3}{RGB}{0,158,115}
\definecolor{c2}{RGB}{213,94,0}
\definecolor{c1}{RGB}{0,114,178}

\usepackage{natbib}
\usepackage{graphicx}

\newcommand\dd{{\partial }}

\title{Generation of zonal flows in convective systems by travelling thermal waves}
\author{Philipp Reiter, Olga Shishkina}
\author{Philipp Reiter\aff{1}
\corresp{\email{philipp.reiter@ds.mpg.de}}, 
Xuan Zhang\aff{1},
Rodion Stepanov\aff{2,3},
\and Olga Shishkina\aff{1}
\corresp{\email{olga.shishkina@ds.mpg.de}}}

\affiliation{\aff{1}Max Planck Institute for Dynamics and Self-Organization,
G{\"o}ttingen, 37075, Germany
\aff{2}Institute of Continuous Media Mechanics, Russian Academy of Science, Perm, 614013, Russia
\aff{3}Perm National Research Polytechnic University, Perm, 614990, Russia
}
\begin{document}
\maketitle

\begin{abstract}
This work addresses the effect of travelling thermal waves applied at the fluid layer surface, on the formation of global flow structures in 2D and 3D convective systems. For a broad range of Rayleigh numbers ($10^3\leq Ra \leq 10^7$) and thermal wave frequencies ($10^{-4}\leq \Omega \leq 10^{0}$), we investigate flows with and without imposed mean temperature gradients. Our results confirm that the travelling  thermal waves can cause zonal flows, i.e. strong mean horizontal flows. We show that the zonal flows in diffusion dominated regimes are driven purely by the Reynolds stresses and end up always travelling retrograde. In convection dominated regimes, however, mean flow advection, caused by tilted convection cells, becomes dominant. This generally leads to prograde directed mean zonal flows. By means of direct numerical simulations we validate theoretical predictions made for the diffusion dominated regime. Furthermore, we make use of the linear stability analysis and explain the existence of the tilted convection cell mode. Our extensive 3D simulations support the results for 2D flows and thus provide further evidence for the relevance of the findings for geophysical and astrophysical systems. 
\end{abstract}


\section{Introduction} 
\label{sec:intro}
The problem of the generation of a mean (zonal) flow in a fluid layer due to a moving heat source is an old one. \cite{Halley1687} was probably the first who perceived that the periodic heating of the Earth's surface, due to Earth's rotation, could be the reason for the occurrence of zonal winds in the atmosphere. Nearly three centuries later, experiments by \cite{Fultz1959}, in which a bunsen flame was rotated around a cylinder filled with water, verified Halley's hypothesis. The moving flame caused zonal flows and the fluid started to move opposite to the direction of the flame. Since then, several experimental and theoretical studies appeared, which illuminated this phenomena.

Thus, \cite{Stern1959} repeated Fultz's experiments using a cylindrical annulus. His observations confirmed the previous results that the fluid acquires a net vertical angular momentum through the rotation of a flame, this time despite the suppression of radial currents in such a domain. Stern then provided a simple two-dimensional (2D) model, showing that the mean motion is maintained through the presence of the Reynolds stresses. \cite{Davey1967} extended Stern's model and provided theoretical explanation that in an enclosed domain, diffusion dominated flows always acquire a net vertical angular momentum in a direction opposite to the rotation of the heat source. His model provided asymptotic scalings for the dependency of the time and space averaged mean horizontal velocity, $\langle U_x\rangle_V$, with the characteristic frequency of the moving heat source $\Omega$: $\langle U_x\rangle_V\sim \Omega^1$ for $\Omega \rightarrow 0$ and $\langle U_x\rangle_V\sim \Omega^{-4}$ for $\Omega \rightarrow \infty$. The topic gained further attention when \cite{Schubert1969} suggested that the 4-day retrograde rotation of the Venus atmosphere might be driven by such a periodic thermal forcing. By using a low Prandtl number ($Pr$) fluid, they observed that the induced mean flow rotated rapidly and exceeded the rotation speed of the heat source, which was rotated below a cylindrical annulus filled with mercury ($Pr\ll 1$), by up to 4 times. This validated the linear analysis by Davey, who predicted the speed of the fluid to increase as $Pr$ becomes small. However, at this time it became clear that the induced rapid mean flows may exceed the range of validity of Davey's linear theory. Consequently, \cite{Whitehead1972}, \cite{Young1972} and \cite{Hinch1971} studied the influence of weakly non-linear contributions. They concluded that the small higher order corrections rather tend to suppress the induced retrograde zonal flows and that the occurring secondary rolls transport momentum in the direction of the moving heat source. It therefore seemed unlikely that the mean flows become much faster than the heat source phase speed, even for small $Pr$, as soon as convective processes come into play.

The preceding analysis certainly lacked the complexity of convective flows, and therefore \cite{Malkus1970}, \cite{Davey1967} and other authors anticipated that convective and shear instabilities could alter the entire character of the solution. In particular, \cite{Thompson1970} showed that the interaction of a mean shear with convection can lead to a tilt of the convection rolls and thus to the transport of the momentum along the shear gradient and thereby amplifies the mean shear flow. In this scenario, the convective flow is unstable to the mean zonal flow even in the absence of a modulated travelling temperature variation, which suggests that the mean zonal flows might be the rule and not the exception to periodic flows that are thermally or mechanically driven. However, the direction of this mean zonal flow would be solely determined by a spontaneous break of symmetry; it could either move counter (retrograde) to the  imposed travelling wave (TW) or in the same directions as the TW (prograde). 

The existence of mean flow instabilities in internally heated convection and in rotating Rayleigh--B{\'e}nard convection (RBC) \citep{Ahlers2009} was studied theoretically by \cite{Busse1972, Busse1983} and \cite{Howard1984}, but has not been observed in laboratory experiments. In classical RBC, a zonal flow, if imposed as an initial flow, can survive \citep{Goluskin2014} but only if the ratio of the horizontal to vertical extensions of the domain is smaller than a certain value, see \cite{Wang2020} and \cite{Wang2020a}. Also, several studies examined the effects of time-dependent sinusoidal perturbations in RBC. \cite{Venezian1969} showed that the onset of convection can be advanced or delayed by modulation, while \cite{Yang2020} and \cite{Niemela2008} demonstrated a strong increase of the global transport properties in some cases. 

Its general nature makes the travelling thermal wave problem appealing to study, however, to our knowledge, there are only a few studies recently published, that are related to the original "moving flame" problem. Therefore in the present study we revisit the existing theoretical models, specifically Davey's model and validate it by means of state of the art direct numerical simulations (DNS). Furthermore, we study a setup with non-vanishing vertical mean temperature gradient (as in RBC), to study the influence of the travelling thermal wave on convection dominated flows and discuss the absolute strength and the direction of the induced zonal flows. Despite the substantial advances over the years, it remains unanswered, whether the thermal travelling wave problem is merely of academical interest or, indeed, of practical relevance in the generation of geo- and astrophysical zonal flows \citep{Maximenko2005,Nadiga2006,Yano2003}. For this purpose, in chapter 2, we complement our analysis with thorough 3D DNS. For the sake of generality, we choose a classical RBC setup. Ultimately, we analyse the absolute angular momentum in 3D flows (respectively, horizontal velocity in 2D flows) and provide insight into the mean flow structures.

\section{Methods}
\label{sec:A}
\subsection{Direct numerical simulations}

The governing equations in the Oberbeck--Boussinessq approximation for the dimensionless velocity ${\bf u}$, temperature $\theta$ and pressure $p$ read as follows: 
\begin{eqnarray*}
\label{eq:1}
   \label{eq:2}
   \dd{\bf u}/\dd t+{\bf u}\cdot \nabla {\bf u}+\nabla {p}&=&
    \sqrt{Pr/Ra} \nabla^2 {\bf u}+ {\theta}{\bf e}_z ,\\
\label{eq:3}
    \dd{\theta}/\dd t+{\bf u}\cdot \nabla {\theta}&=&1/\sqrt{Pr Ra} \nabla^2 {\theta}, \quad \nabla \cdot {\bf u} =0.
\end{eqnarray*}
Here $t$ denotes time and ${\bf e}_z$ the unit vector in the vertical direction. The equations have been non-dimensionalised using the free-fall velocity $u_{ff}\equiv (\alpha g \Delta \hat{H})^{1/2}$, the free-fall time $t_{ff}\equiv \hat{H}/u_{ff}$, $\Delta$ the amplitude of the thermal TW and $\hat{H}$ the cell height. The dimensionless parameters $Ra$, $Pr$ and the aspect ratio $\Gamma$ are defined by:
\begin{eqnarray*}
\qquad \quad
Ra\equiv \alpha g \Delta \hat{H}^3/(\kappa\nu), \qquad Pr\equiv\nu/\kappa, \qquad
\Gamma\equiv \hat{L}/\hat{H},
\end{eqnarray*}
where $\hat{L}$ is the length of the domain, $\nu$ is the kinematic viscosity, $\alpha$ the isobaric thermal expansion coefficient, $\kappa$ the thermal diffusivity and $g$ the acceleration due to gravity. This set of equations is solved using the finite-volume code \textit{goldfish} \citep{Kooij2018,Shishkina2015}, which employs a fourth-order discretization scheme in space and a third order Runge--Kutta, or, alternatively, an Euler-Leapfrog scheme in time. The code runs on rectangular and cylindrical domains and has been advanced for a 2D-pencil decomposition for a highly parallel usage. The spatial grid resolution of the simulations was chosen according to the  minimum resolution requirements of \cite{Shishkina2010}. A stationary state is ensured by monitoring the volume-averaged, the wall-averaged and the kinetic dissipation based Nusselt numbers.

In this study the following notations are used: Temporal averages are indicated by an overline or by a capital letter, thus the Reynolds decomposition of the velocity reads $u=U+u^\prime$, decomposing $u$ into its mean part $U$ and fluctuation part $u^\prime$. Unless specifically stated, time averages are carried out over a long period of time, however, in section \ref{sec:2d-prograde}, the averaging period was deliberately restricted to only a few wave periods to achieve a time scale separation. Further, the spatial averages are denoted by angular brackets $\langle \cdot \rangle$, followed by the respective direction of the average, e.g. $\langle \cdot \rangle_x$ denotes an average in $x$; $\langle \cdot \rangle_V$ denotes a volume average. And ultimately, the velocity vector definitions $\mathbf{u} \equiv (u_x,u_y,u_z) \equiv (u,v,w)$ are used interchangeably. 

\subsection{Theoretical model}
\label{sec:model}
Already the earliest models proposed by \cite{Stern1959} and \cite{Davey1967} gave a considerable good understanding of the moving heat source problem. Although there are more complex models \citep{Stern1971} based on adding higher order non-linear contributions \citep{Whitehead1972,Young1972,Hinch1971,Busse1972}, this section focuses on revisiting the main arguments of Davey's original work, which is expected to give reasonably good results in the limit of small $Ra$. Besides, a more complete derivation and concrete analytical solutions are provided in Appendix \ref{app:theory}.

Given the linearized Navier--Stokes equations in two dimensions and averaging the horizontal momentum equation in the periodic $x$--direction and over time $t$, one can derive the following balance:
\begin{align}
\sqrt{Pr/Ra} {\partial_z^2} \langle U \rangle_x = \partial_z \langle \overline{u^\prime w^\prime} \rangle_x + \langle W \partial_z U \rangle_x.
\label{eq:balance}
\end{align}
Evidently, a mean zonal flow $\langle U \rangle_x$ is maintained by the momentum transport due to the Reynolds stress component $\overline{u^\prime w^\prime}$ and by mean advection through $W \partial_z U$. The theory further advances by assuming that no vertical mean flow exists ($W=0$), which reduces equation (\ref{eq:balance}) to the balance between viscous mean diffusion and Reynolds stress diffusion. Furthermore, by neglecting convection and variations in $x$, the linearized equations can be written as a set of ordinary differential equations, that can be solved sequentially to find $u^\prime$ and $w^\prime$ and ultimately the Reynolds stress term $\overline{u^\prime w^\prime}$. This procedure is shown in Appendix \ref{app:theory}. Given the Reynolds stress field, equation (\ref{eq:balance}) has to be integrated twice to obtain the mean zonal flow $U(z)$. Integrating that profile again finally gives the total mean zonal flow $\langle U \rangle_V$, which is an important measure of the amount of horizontal momentum or, respectively, angular momentum in cylindrical systems, that is generated due to the moving heat source. The last step can be solved numerically, however, following \cite{Davey1967}, the limiting relations can be calculated explicitly: 

\begin{align}
&\langle U_x \rangle_V  = -\frac{\pi}{2}\frac{ k^3 Ra^2 Pr^{-2}(Pr+1)}{ 12!}  \Omega + \mathcal{O}(\Omega^3)  \ \text{for}\ \Omega \rightarrow 0,\\
\begin{split}
&\langle U_x \rangle_V = -\frac{k^3 Ra^{-1/2} Pr^{-3/2}}{ 256 \pi^4 (Pr+1)} \Omega^{-4} + \mathcal{O}(\Omega^{-9/2})  \ \text{for}\ \Omega \rightarrow \infty,
\end{split}
\label{eq:limit}
\end{align}
where the horizontally travelling wave, $\theta(x,t)=0.5 \cos(kx-2\pi \Omega t)$, is applied to the bottom and top plate. We would like to add that this theoretical model is, as determined by its assumptions, expected to be limited to diffusion dominated, small--$Ra$ flows. However, when momentum and thermal advection take over, its validity remains questionable. We will show later that after the onset of convection, where eventually mean advection takes over, the neglect of the $W \partial_z U$--contribution is no longer justified. 

\begin{figure}
\begin{minipage}[c]{0.999\textwidth}
\centering
\includegraphics[width=0.8\textwidth]{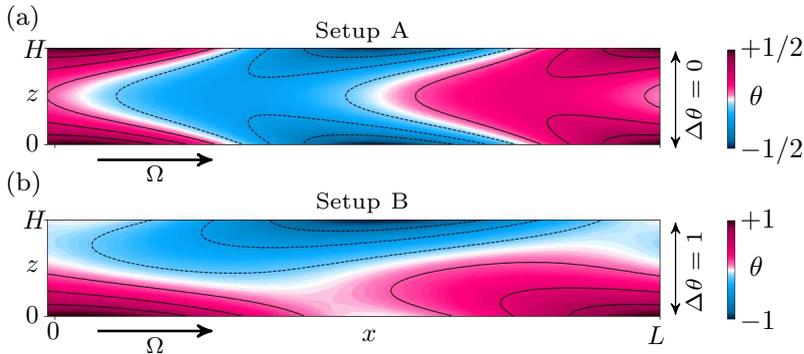}
\end{minipage}
\caption{Sketch of the 2D numerical setup. The colour represents the dimensionless temperature distribution for the purely conductive cases ($\Omega=0.1$). The thermal wave is imposed at the top and bottom plates, propagating to the right, in the positive $x$-direction. (a) Setup A: no mean temperature gradient is imposed between the top and the bottom. (b) Setup B: with (unstably stratified) mean temperature gradient, like in RBC. The figure shows temperature snapshots, while the time averaged conduction temperature field depends linearly on $z$. }
\label{fig:2d-setup}
\end{figure}

\section{2D-convective system}

As described by \cite{Stern1959}, the generation of a laminar zonal flow by a TW can be successfully explained in a 2D system, which makes it a good starting point. The temperature boundary conditions (BCs) are time- and space-dependent,
\begin{align*}
\theta(x,z=0,t) &=  0.5\left[\cos(x-2\pi\Omega t) + \Delta \theta \right], \\
\theta(x,z=H,t) &=  0.5\left[\cos(x-2\pi\Omega t) - \Delta \theta \right].
\end{align*}
Here, $\Omega$ indicates the temporal frequency of the travelling TW in free-fall time units. For example, $\Omega=10^{-1}$ describes a wave with a period of $10$ free-fall time units $\tau_{ff}$, and $\Delta \theta$ is introduced as a control parameter for the strength of the mean temperature gradient.

In the following, two different setups are considered. In setup A (figure \ref{fig:2d-setup} a) -- the one originally examined by \cite{Davey1967} -- no mean temperature gradient exists ($\Delta \theta=0$) and the top and bottom plate temperatures are equal, whereas in setup B (figure \ref{fig:2d-setup} b) a mean, unstable temperature gradient is applied ($\Delta \theta=1$). For simplicity, the mean temperature gradient is set equal to the amplitude of the thermal wave. In this setup, effects of convection are expected to become dominant. Averaged over time, this setup resembles RBC, therefore, it can be regarded as a spatially and temporally modulated variant of RBC. Further, no-slip conditions are applied at the top and bottom plates, the $x$-direction is periodic and the domain has the length $L=2\pi$ and the height $H=1$. In upcoming studies, one might introduce a second Rayleigh number based on the mean temperature gradient (as in RBC), namely $Ra_{\Delta \theta} \equiv \alpha g \Delta \theta \hat{H}^3/(\kappa\nu)$. However, in this work the connection to $Ra$ is simply $Ra_{\Delta \theta}=0$ for Setup A and $Ra_{\Delta \theta}=Ra$ for Setup B.

\begin{figure}
\begin{minipage}[c]{0.99\textwidth}\centering
\includegraphics[width=0.99\textwidth,trim=0 0 0 0,clip]{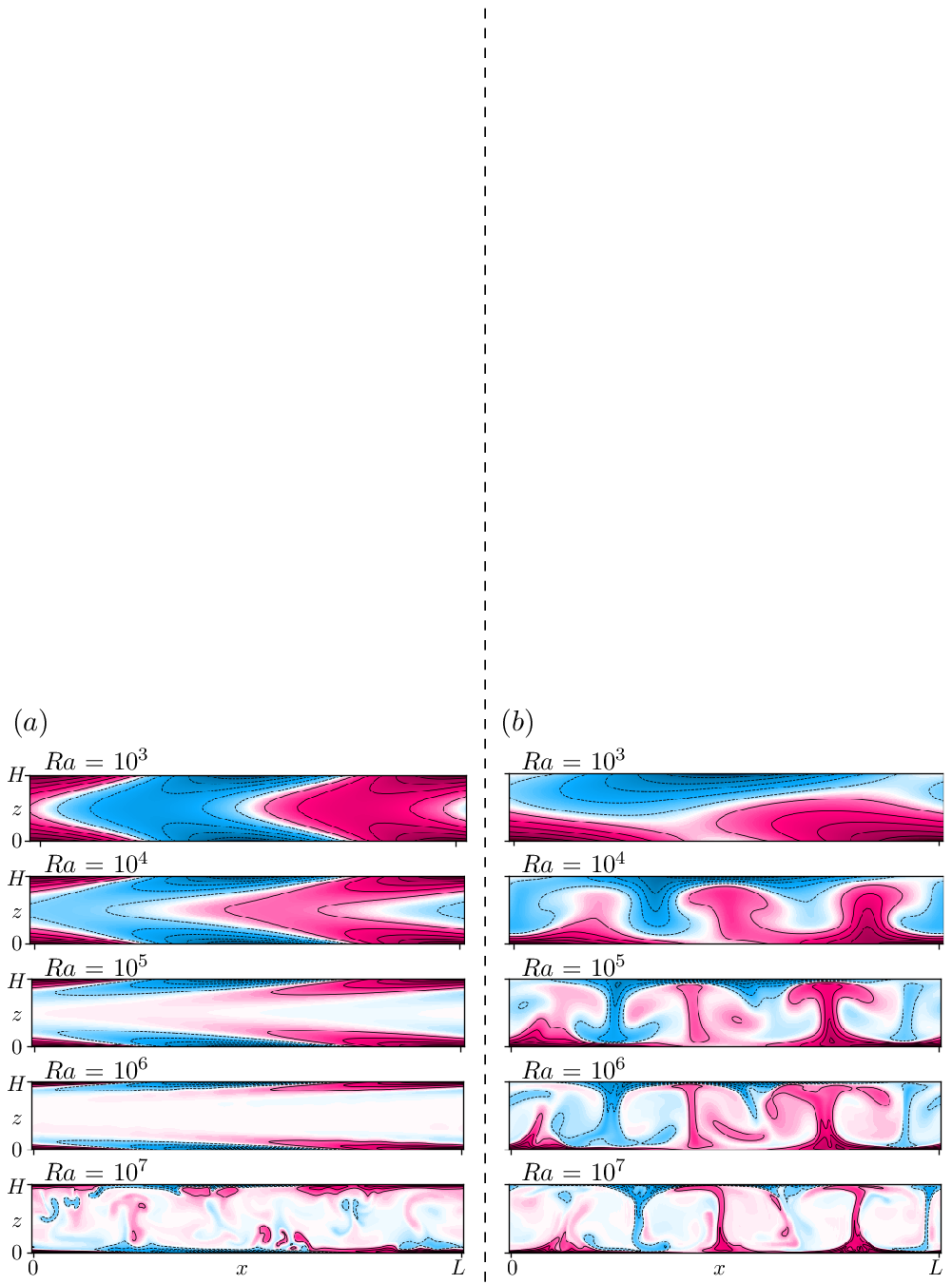}
\end{minipage}
\caption{ Snapshots of the temperature field $\theta$ at a fixed TW speed $\Omega=0.1$ (propagating to the right). (a) Setup A and (b) Setup B. The plumes in Setup B travel either retrograde or prograde (see supplementary movies).} 
\label{fig:2d-snapshot}
\end{figure}

The overall focus in this study lies on variations of the zonal flow with $Ra$ and $\Omega$. Thus, the parameter space spans $10^3 \leq Ra \leq 10^7$ and $10^{-4}\leq \Omega \leq 10^0$, while the aspect ratio and Prandtl number are kept constant ($\Gamma=2\pi$, $Pr=1$). Exemplary temperature fields at a fixed $\Omega=0.1$ are shown in figure \ref{fig:2d-snapshot}.

\subsection{Results}
\label{sec:2d-results}

\begin{figure}
\begin{minipage}[c]{0.49\textwidth}
\centering
\includegraphics[width=0.99\textwidth]{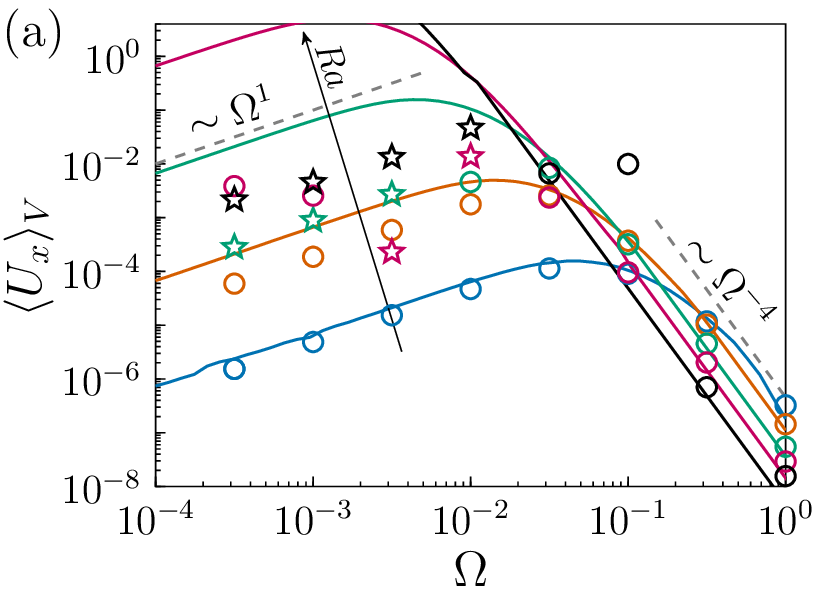}
\end{minipage}
\hfill
\begin{minipage}[c]{0.49\textwidth}
\centering
\includegraphics[width=0.99\textwidth]{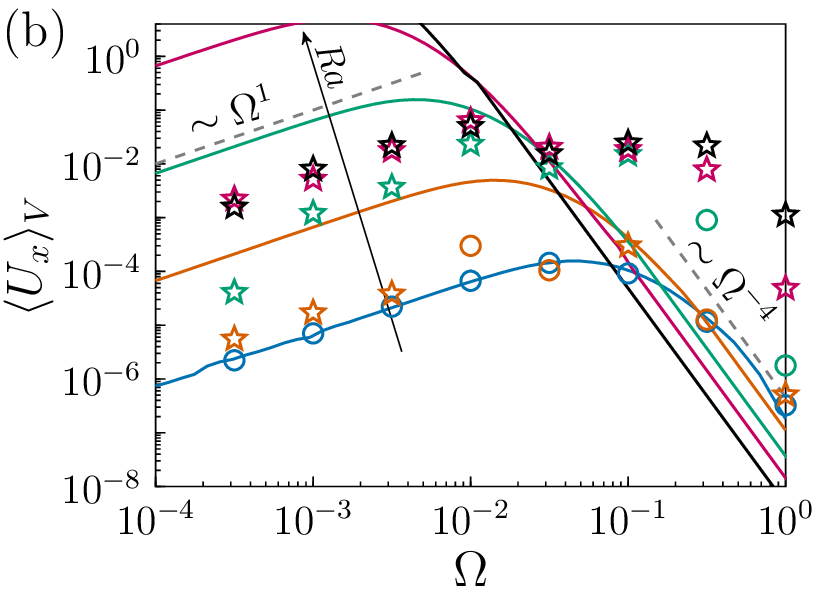}
\end{minipage}

\caption{Mean velocity of the zonal flow vs. the wave frequency $\Omega$ for $Ra =$ $10^3$ (blue), $10^4$ (orange), $10^5$ (green), $10^6$ (red) and $10^7$ (black). Circles (stars) denote a retrograde (prograde) mean zonal flow, the solid lines of the corresponding colour show the results of the theoretical model by \cite{Davey1967}. (a) Setup A and (b) Setup B. }
\label{fig:2d_UvsOm}
\end{figure}

The theoretical model, as presented in Appendix \ref{app:theory}, aims to explain the generation of the total mean momentum $\langle U_x \rangle_V$ for a given $Ra$ and wave frequency $\Omega$. Moreover, it predicts that the generated mean momentum will be directed opposite, i.e. retrograde, to the travelling thermal wave. In this section we study the validity of the model and reveal its limitations.

Figure \ref{fig:2d_UvsOm} shows the numerical data from the DNS together with the respective results of the theoretical model, for different $Ra$. Worth noting first is, that the maximum of the theoretical model is located at a fixed frequency, if the frequency is expressed in terms of the diffusive timescale rather than the free-fall time scale $\Omega_{\kappa,max} = \Omega/\sqrt{RaPr}  \approx 0.66$. This indicates, that the model predictions could be collapsed onto a single curve. Nonetheless, this was avoided here for the sake of clarity. 

We begin our discussion with the results of setup A, shown in figure \ref{fig:2d_UvsOm} (a). The theoretical model by \cite{Davey1967}, indicated by the solid lines, gives accurate results for $Ra=10^3$ and a good agreement for $Ra=10^4$, although, evidently the model systematically overestimates the mean momentum generation for higher $Ra$. In fact, this is consistent with \cite{Whitehead1972}, \cite{Young1972} or \cite{Hinch1971} who observed that corrections of higher order non-linear contributions tend to suppress the induced retrograde zonal flows. Also it suggests that an induced mean flow does not strengthen itself, i.e. there is no positive feedback mechanism between the mean flow and Reynolds stresses. While all low $Ra$ flows and high $Ra$ flows in the limit of large $\Omega$ are well predicted by the model, the large $Ra$ flows are mostly over predicted (except $Ra=10^7$ and $\Omega=0.1$, the only flow of that setup that becomes truly turbulent, despite similar initial conditions). Presumably even more important is that some of the flows for $Ra \geq 10^5$ exhibit a positive/prograde mean flow, indicated by a star symbol, which is especially prevalent at small $\Omega$.

\begin{figure}
\begin{minipage}[c]{0.449\textwidth}
\centering
\hskip8mm
\includegraphics[width=0.97\textwidth]{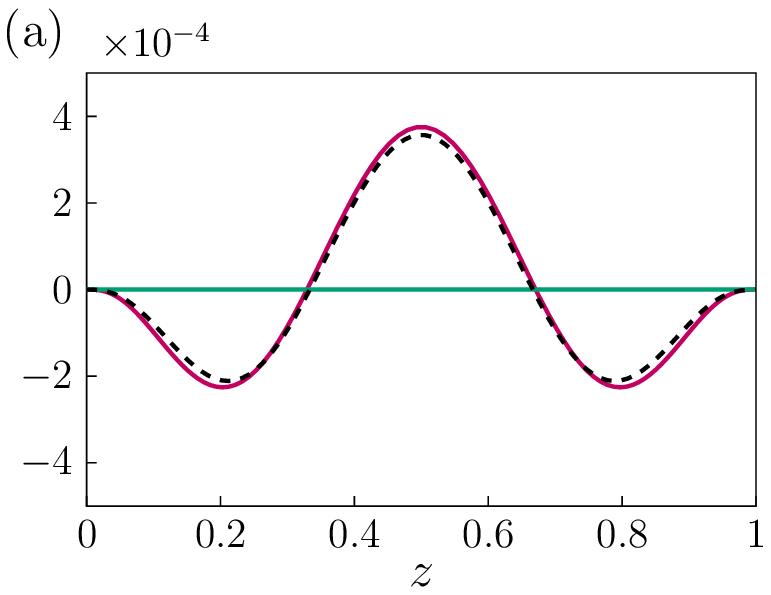}
\end{minipage}
\begin{minipage}[c]{0.12\textwidth}
\centering
\includegraphics[width=0.97\textwidth,trim=0 0 0 0,clip]{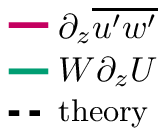}
\end{minipage}
\hfill
\begin{minipage}[c]{0.449\textwidth}
\centering
\hskip8mm
\includegraphics[width=0.97\textwidth,trim=0 0 0 0,clip]{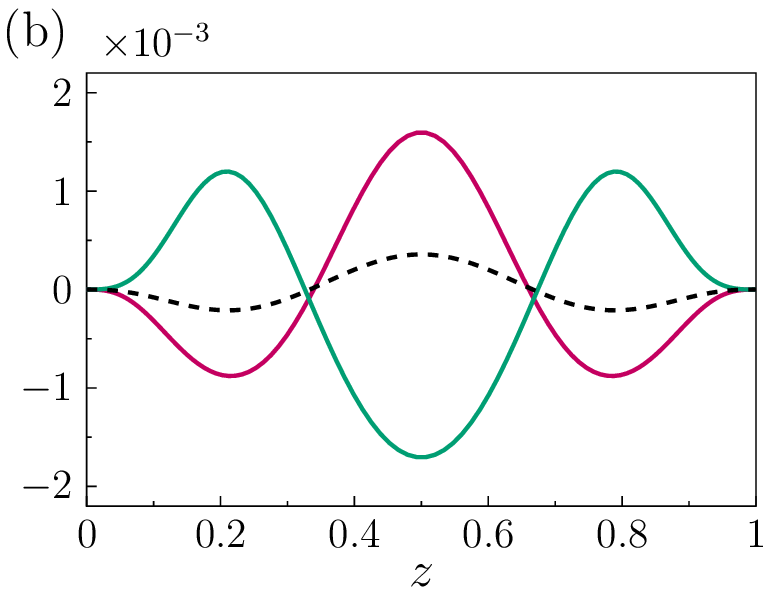}
\end{minipage}
\caption{Mean profiles of Reynolds stress vs. mean flow advection contribution for $Ra=10^4$ and $\Omega=0.1$ for (a) ${\Delta \theta=0}$ and (b) ${\Delta \theta=1}$, where mean advection dominates. The flow in (a) moves retrograde due to the Reynolds stress contribution, while the flow in (b) shows a prograde mean flow ($\sqrt{Pr/Ra} {\partial_z^2} \langle U \rangle_x = \partial_z \langle \overline{u^\prime w^\prime} \rangle_x + \langle W \partial_z U \rangle_x$).} 
\label{fig:2d_balance}
\end{figure}

Turning the focus to Setup B, shown in figure \ref{fig:2d_UvsOm} (b), the differences become even more obvious, since adding a mean temperature gradient enhances the effects of convection further. For $Ra = 10^3$ the picture is clear, as it is below the onset of convection $Ra_c\approx 1708$ for classical RBC even for the unbounded domains. The Reynolds stresses remain dominant, which preserves the development of a mean flow opposite to the TW direction. However, for $Ra \ge 10^3$ all but a few of the simulations end up with a prograde mean flow final state. In order to understand the role of the mean flow, we analyse the two terms on the right side of equation (\ref{eq:balance}), which are presented in figure \ref{fig:2d_balance}. The model neglects mean advection, it only captures contributions of $\overline{u^\prime w^\prime}$. As seen in figure  \ref{fig:2d_balance} (a), this is justified for a flow without strong convection effects and the model predictions agree well with the Reynolds stresses obtained in the simulations. This is different from the situation in figure \ref{fig:2d_balance} (b), where obviously mean flow advection $W \partial_z U$  starts to take over. The shape of the mean flow advection curve is antiphase to the Reynolds stress curve and contributes the most. This explains the reversal of the mean flow, from retrograde in figure \ref{fig:2d_balance} (a) to prograde in figure \ref{fig:2d_balance} (b). 

The underlying reason for that will be examined in more detail in the next section. But briefly, the main argument is that there exist two competing mechanisms, one induced by the TW and the other induced by convection rolls, which act on different time scales. At small $Ra$, as convection rolls move considerably slower, an average over a few TW time periods can reliably separate both structures, so that the Reynolds stresses reflect mainly the TW contributions, while the mean field represents the convection rolls. Therefore, the dominant mean flow advection in figure \ref{fig:2d_balance} (b) reflects the dominance of advection by convection rolls as $Ra$ increases.

A few more interesting observations can be deduced from figure \ref{fig:2d_UvsOm} (b). First, compared to the theory, the simulations show significantly larger values at high $Ra$. Apparently, the mean zonal flow can be substantially stronger than expected and its velocity can exceed the TW phase velocity. Second, while the theory predicts the location of the maximum zonal flow at a constant diffusive time scale, the DNS indicates a coupling with the free-fall time rather than with the diffusive time and the maximum is found in the region $0.01\leq \Omega \leq 0.1$. This is in so far important, since natural flows often fall within this parameter range. We show this in the context of the Earth's atmosphere in section \ref{sec:3d}. Finally the instantaneous fields most often show three plumes (figure \ref{fig:2d-snapshot} b), while a classical RBC simulation with the same initial conditions would develop four plumes. Presumably, either the sinusoidal temperature distribution at the plates, or a preexisting shear flow (before Rayleigh--Taylor instabilities develop) reduces the number of plumes. On this basis, we tested the linear stability of the Rayleigh--Taylor instabilities with an imposed shear flow, and found indeed that the wavelength of the most unstable mode decreases.

\begin{figure}
\begin{minipage}[c]{0.499\textwidth}
\centering
\hskip8mm
\includegraphics[width=0.99\textwidth]{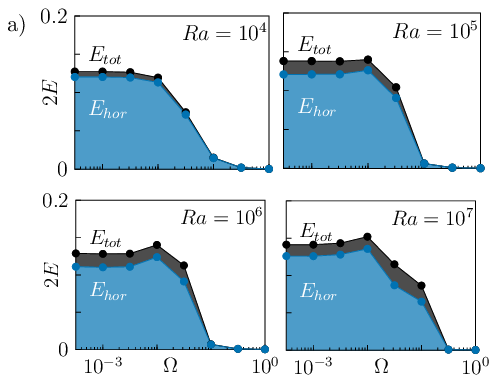}
\end{minipage}
\begin{minipage}[c]{0.499\textwidth}
\centering
\includegraphics[width=0.99\textwidth]{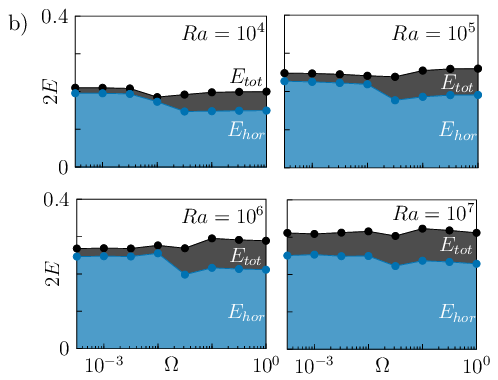}
\end{minipage}
\caption{Total kinetic energy $E_{tot}$ (black) and horizontal kinetic energy $E_{hor}$ (blue) for (a) Setup A and (b) Setup B. }
\label{fig:2d_energy}
\end{figure}

 In figure \ref{fig:2d_energy} we show the total kinetic $E_{tot}=\frac{1}{2}\langle \overline{u_x^2+u_z^2} \rangle_V^{1/2}$ and horizontal (zonal flow) kinetic energy $E_{hor}=\frac{1}{2}\langle \overline{u_x^2}\rangle_V^{1/2}$ in order to elucidate the energetic impact of the present zonal flows and to evaluate the strength of the vertical and horizontal motions. Setup A (a) is clearly dominated by the horizontal kinetic energy throughout the whole parameter range.  For $\Omega>10^{-1}$, the kinetic energy drops close to zero and the temperature is transported by conduction only above this limit. However, before the kinetic energy drops, the curves show an energy enhancement. The location of the energy maximum coincides with the maximum of the zonal flow (figure \ref{fig:2d_UvsOm} a), which indicates that the zonal flows can have a significant imprint in the energy of the system.  Likewise, setup B (figure  \ref{fig:2d_energy} b) is also dominated by horizontal kinetic energy. Though, obviously for larger $Ra$ and larger $\Omega$, the magnitude of the vertical kinetic energy becomes increasingly important. This further supports that the neglect of the vertical velocity component $W$ is eventually no longer justified for these parameter regimes.

\subsubsection{Origin of prograde flows in convection dominated flows}
\label{sec:2d-prograde}

\begin{figure}
\centering
\begin{minipage}[c]{0.45\textwidth}
\centering
\includegraphics[width=0.99\textwidth,trim=0 0 0 0,clip]{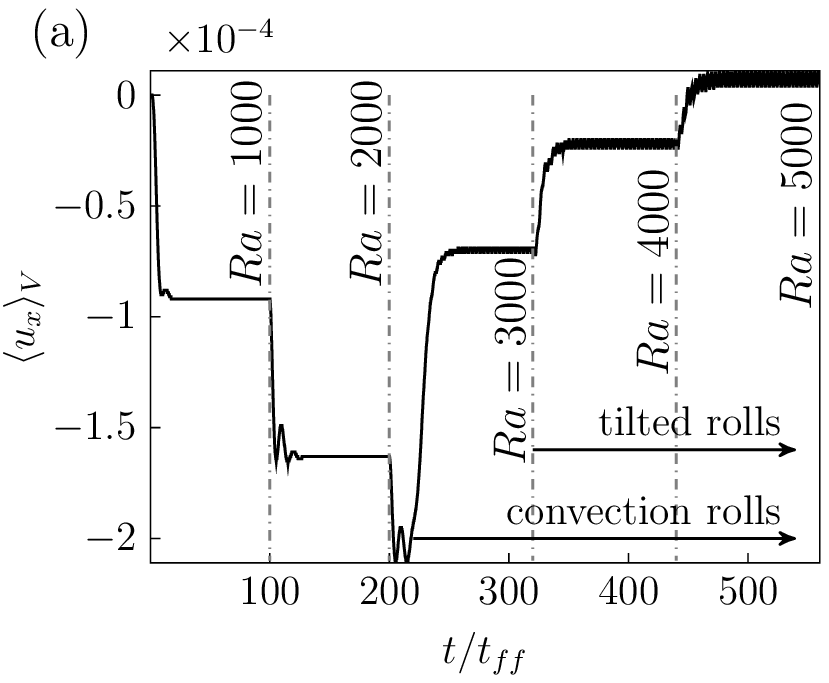}
\end{minipage}
\begin{minipage}[c]{0.40\textwidth}
\centering
\includegraphics[width=0.99\textwidth,trim=0 0 0 0,clip]{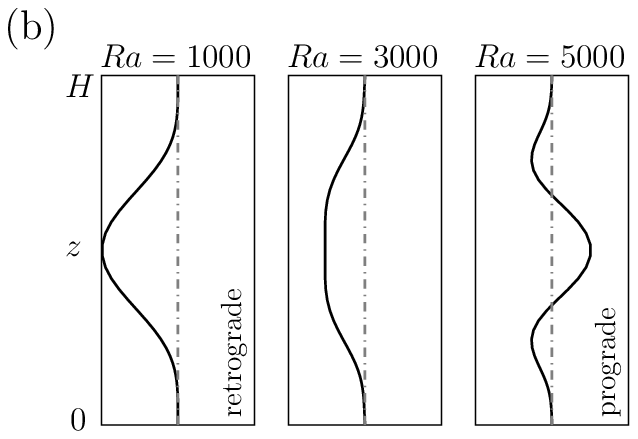}
\end{minipage}

\centering
\begin{minipage}[c]{0.88\textwidth}
\centering
\includegraphics[width=0.99\textwidth,trim=0 0 0 0,clip]{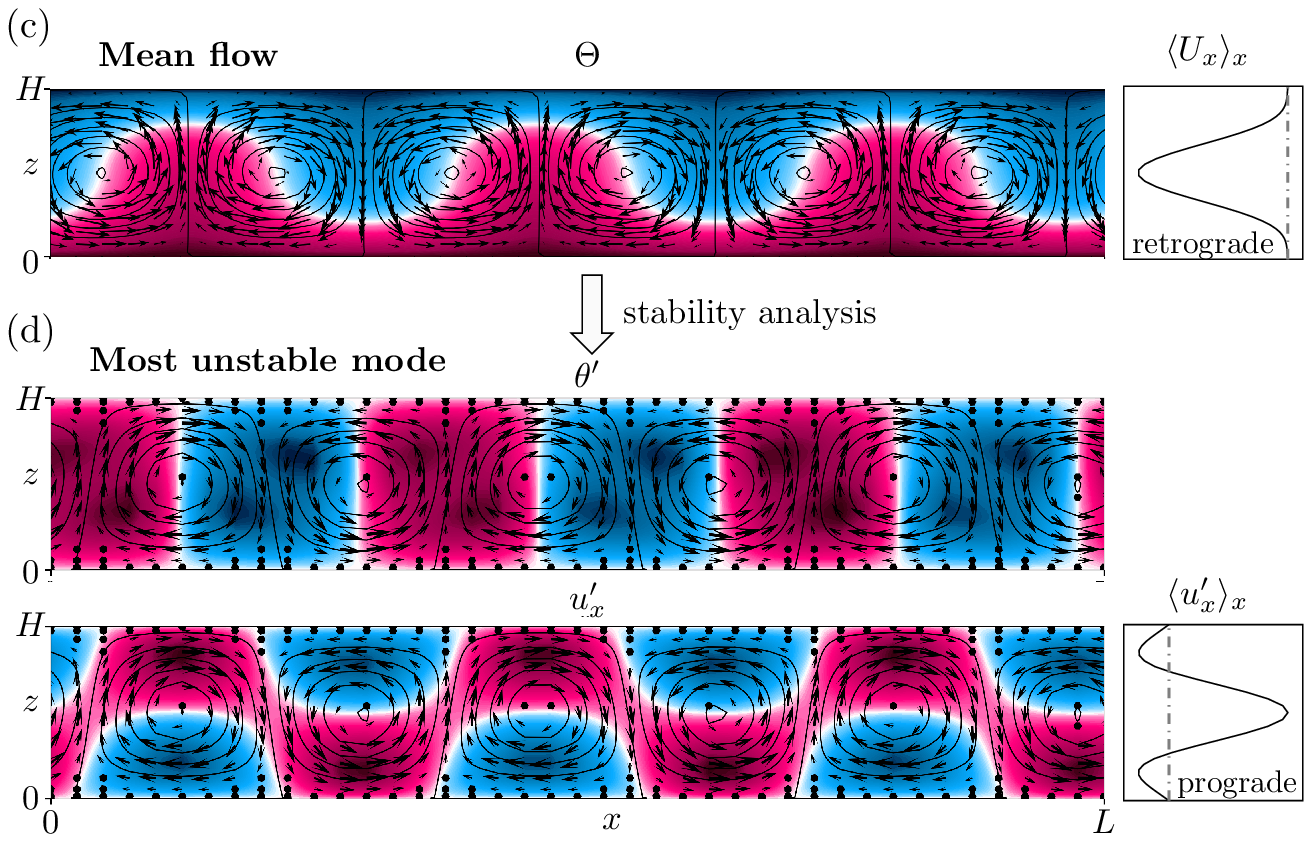}
\end{minipage}
\caption{Path from a retrograde flow to a prograde flow. (a) Time evolution of the mean zonal flow; $Ra$ was increased step-wise. For $Ra\geq 3000$ convection rolls form; for $Ra\geq 4000$ the rolls tilt significantly, and the mean zonal flow becomes positive. (b) $u_x$ profiles for $Ra=1000, 3000, 5000$. (c) Mean flow extracted at $Ra=3000$ (averaged over one TW period). (d) Result of the global stability analysis for the mean flow of (c), that becomes unstable for $Ra\geq 4000$ to tilted convection rolls.  }
\label{fig:2d_stability}
\end{figure}

In order to understand how prograde flows can emerge, we looked at the route from small to large $Ra$ for a specific configuration. Setup B and $\Omega=0.1$ is well suited for this purpose, since the transition from a retrograde flow to a prograde flow appears early, already below $Ra=10^4$ (figure \ref{fig:2d_UvsOm} b). Thus, a simulation was initiated at $Ra=1000$ and then $Ra$ was progressively increased by $1000$ each time after a steady state has settled. The time evolution of the total mean zonal flow is given in figure \ref{fig:2d_stability} (a). At the lowest $Ra$, the mean flow is retrograde. Increasing $Ra$ to $2000$ enhances its strength further, as anticipated. But already at $Ra=3000$ the zonal flow breaks down and its vertical profile, as seen in figure \ref{fig:2d_stability} (b), flattens. Ultimately, at $Ra\geq 4000$ this profile flips over and the total zonal flow turns into a prograde state. 

As we have shown in the preceding analysis (figure \ref{fig:2d_balance} b), in the presence of convection cells, the mean zonal flow can be fed by the base flow itself, in particular it is fed by the vertical advection of horizontal momentum $W\partial_z U$. Now, let us consider perfectly symmetric convection cells;  although locally, at a position in $x$, momentum may be transported up- or downward, the symmetry, however, would balance this transport at another location and the net transport would become zero. Therefore there must be a symmetry breaking in the convection cells, which correlates $W$ with $\partial_z U$. A possible mechanism, even discussed in the context of the moving heat source problem, was described by \cite{Thompson1970} and theoretically analysed by \cite{Busse1972}, who showed that, in a periodic domain, convection rolls can become unstable to a mean shear flow. This mean shear tilts the convection cells such that their asymmetric circulation maintains a shear flow. In the following this mean flow instability will be called tilted cell instability.
\cite{Busse1972} showed the existence of this instability on a analytic base flow field. Differently, in the following we conduct a stability analysis on a base flow extracted from the DNS.

The first rise of the curve in figure \ref{fig:2d_stability} (a) at $Ra=3000$ coincides with the observed onset of convection, which is slightly delayed compared to classical, unmodulated RBC ($Ra_c\approx 1708$). The convection cells at that point appear to be standing still, almost unaffected from the TW and clearly orders of magnitudes slower than the TW. Therefore a short time average, over one wave period, was applied to separate both timescales, which results in the base flow as shown in figure \ref{fig:2d_stability} (c). Based on this base flow, a linear, temporal stability analysis of the full 2D linearized Navier-Stokes equations was conducted. Details herefore are given in the Appendix \ref{app:lst}. While no unstable mode was detected for $Ra=3000$, for $Ra=4000$ the mean flow becomes unstable, to the mode presented in figure \ref{fig:2d_stability} (d). The growth rate of it is $\sigma \approx 0+0.2i$, suggesting no oscillatory behaviour (real part is zero) but exponential temporal growth (imaginary part larger than zero). This mode shares characteristics with the tilted cell instability described by \cite{Thompson1970}, in the sense that the mode induces a mean shear flow (see profile in figure \ref{fig:2d_stability} d). However, rather than the "pure" shear flows as presented by \cite{Thompson1970} and \cite{Busse1972,Busse1983} with a vanishing total net momentum when integrated vertically, the fluctuation profile found in our study (figure \ref{fig:2d_stability} d on the right) shows a more directed flow, negative in the vicinity of the plates and stronger positive in the center. And especially interesting, its momentum profile has a similar shape as the final state solution of typical prograde flows, e.g. the profile on the right in figure \ref{fig:2d_stability} (b). A few more notes are necessary. The difference between the shape of the mode found in this work, compared to the ones from Thompson and Busse might be explained by different BCs, as both authors applied free-slip conditions at the plates, in contrast to our no-slip conditions. In addition, in their seminal works and in the work of \cite{Krishnamurti1981}, it was already remarked that the mean flow transition is caused by a  spontaneous symmetry breaking and therefore the direction of the shear flow is somewhat arbitrary as it depends on the initial conditions. Indeed, a change in the grid size of the stability analysis led to a most unstable mode with a reversed shear flow profile compared to the mode shown in figure \ref{fig:2d_stability} (d). And finally, even though in figure \ref{fig:2d_stability} (a) tilted rolls are shown to start later as convection rolls, it actually is likely that the convection cells tilt as soon as convection sets in, it is just not clearly visible from the flow fields at that point. 

In a nutshell, the mean flow is unstable -- even in the absence of a boundary temperature modulation -- to a mode with tilted convection cells and non-zero total mean horizontal velocity. Both modes, prograde and retrograde, are found in the global stability analysis, thus it remains unanswered why the DNS at high $Ra$ almost exclusively end up moving in the same direction as the TW. The disturbance velocity profiles resemble those of the final mean flow velocity profiles, therefore, the presented mean flow instability is a plausible mechanism for the generation of moderate strong zonal flows after onset of convection, then dominating over the Reynolds stress mechanism, that is inherent to diffusion dominated flows.

\subsubsection{Space-time structures}
\label{sec:spacetime}

\begin{figure}
\begin{minipage}[c]{0.995\textwidth}
\centering
\includegraphics[width=0.995\textwidth,trim=0 0 0 0,clip]{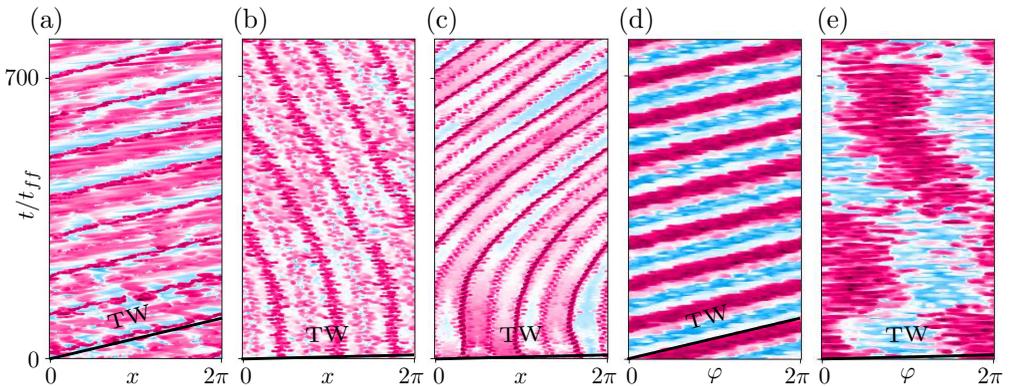}
\end{minipage}
\caption{Evolution of the temperature at mid-height $z=H/2$, for (a-c) 2D flows and (d,e) 3D RBC ($r=0.99R$) flows. (a) $\Omega=0.01$, setup A, (b) $\Omega=0.1$, setup A, (c) $\Omega=0.1$, setup B, (d) $\Omega=0.01$, 3D RBC and (e) $\Omega=0.1$ 3D RBC.  All figures show $Ra=10^7$. The black solid line in each plot indicates the TW speed. }
\label{fig:spacetime}
\end{figure}

The flows found in this study revealed surprisingly rich formations. Therefore this part will be completed with examples of some space-time structures that have been observed in the 2D system and, already ahead of the next part, in the 3D cylindrical system. In addition, movies are provided as supplementary material.

In general, in 2D, as can be seen from figure \ref{fig:2d-snapshot} the temperature field is either symmetric around the horizontal mid-plane (Setup A), or not; in this case there exist plumes (Setup B). In the latter case, there are usually three up- and three down-welling plumes identifiable.  
In the 3D case, the flow consists of rising and falling plumes, which together form a large scale circulation (LSC). If the TW propagates slowly (small $\Omega$), the plumes (2D) or respectively the LSC plane (3D) drift with the same speed as the TW and both structures appear to be connected. However, as $\Omega$ increases and, hypothetically, the TW timescale $\tau_\Omega$ becomes small compared to thermal diffusion $\tau_\kappa$ ($\tau_\kappa/\tau_\Omega = \sqrt{Pr Ra}\Omega$), the plumes (2D) or LSC (3D) "break-off" from the TW, forming two separate structures, acting on different timescales. 

Figure \ref{fig:spacetime} shows the space-time structures of the temperature field, evaluated at mid-height, and in the 3D case at mid-height and near the sidewall. The structures at mid-height either (i) travel with the same speed (but a phase difference) as the thermal wave (a,d), or travel with phase speeds different to the thermal wave and in this case either (ii) retrograde (b,e) or  (iii) prograde (c). Regime (i) is expected for small $Ra$ and/or small $\Omega$ parameters, (ii) is found for large $Ra$ and large $\Omega$, if no mean temperature is present, and (iii) exist in strongly convection dominated flows for large $Ra$ and large $\Omega$, especially if a mean temperature gradient is present. Furthermore it is striking that temperatures between the left and the right region in the vicinity of the plumes center (hottest or coldest regions in figure \ref{fig:spacetime}) do not necessarily fill with the same temperature (c). This gives further evidence for a mean flow instability, as it features similarities of the temperature field of the unstable mode given in figure \ref{fig:2d_stability} (d), due to which a plume loses its horizontal symmetry. Considering the speed of the drifting plumes (b,c), we observe initially exponential growth, as anticipated from an instability, followed by a, possibly, non-linear saturation.

\section{3D-convective systems}

\label{sec:3d}
The preceding part, as most of the existing literature, is confined to 2D flows. Now we will discuss the moving heat source problem in the context of more complicated 3D convective flows. In general, travelling wave solutions are common amongst 3D convective systems. \cite{Bensimon1990,Kolodner1988, Kolodner1988b} observed convection rolls propagating azimuthally in a large aspect ratio annulus near the onset of convection. Their drift velocity is of the order of magnitude $10^{-4} $ to $10^{-3}$, however, drift velocity is not necessarily equal to the mean azimuthal flow. Another kind of travelling wave solution in RBC systems are the spiral patterns found in large aspect ratio cells \citep{Bodenschatz1991,Bodenschatz2000}. These spirals are rotating in either direction, although corotating spirals are more numerous \citep{Cross1995}, and are known to be coupled with an azimuthal mean flow \citep{Decker1994}. Furthermore, in rotating systems travelling wave structures are quite common \citep{Knobloch1990}. These structures are strongly geometry-dependent \citep{Wang2012} and known to induce mean zonal flows that propagate pro- and retrograde \citep{Zhang2020}.

Despite the vast literature on these phenomena, quantitative data on mean flows that are induced by external travelling thermal waves in 3D flows seems to be rare. Therefore our main goal in this part is to gain insight on the strength and structure of such mean flows, and discuss whether their order of magnitude is relevant in natural flows. For this purpose we took the paradigm convective system cylindrical RBC and studied it by means of direct numerical simulations. 

\begin{figure}
\begin{minipage}[c]{0.499\textwidth}
\centering
\includegraphics[height=0.75\textwidth]{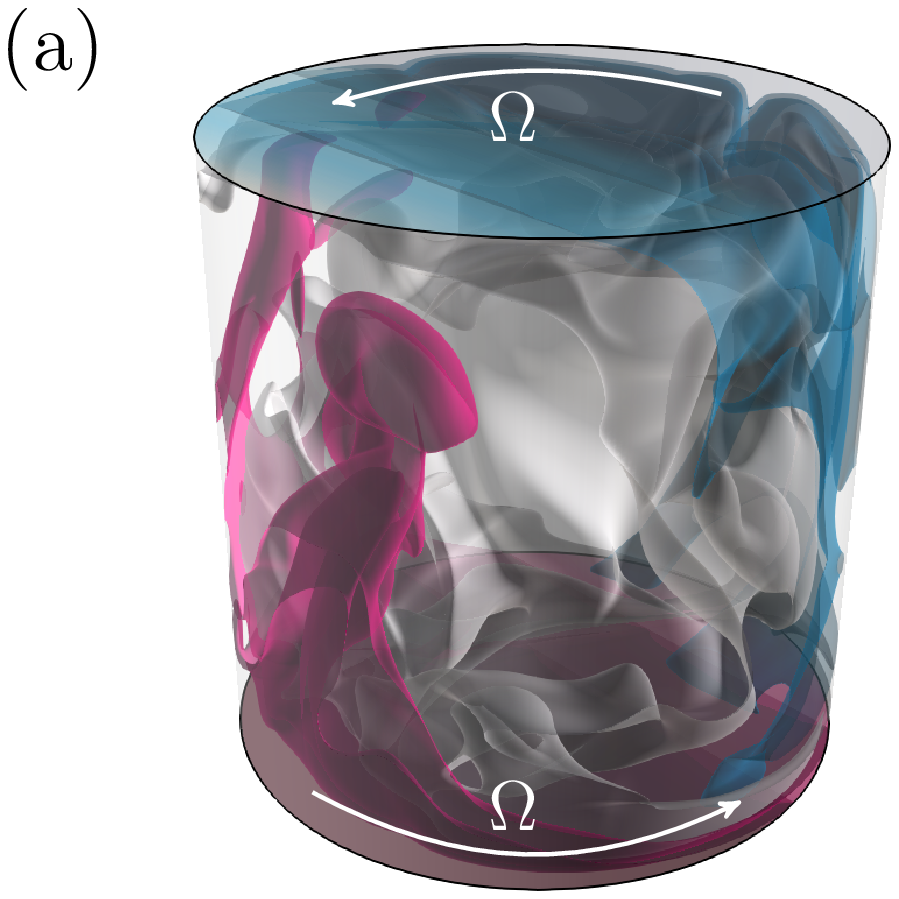}
\end{minipage}
\hfill
\begin{minipage}[c]{0.499\textwidth}
\centering
\includegraphics[height=0.7\textwidth]{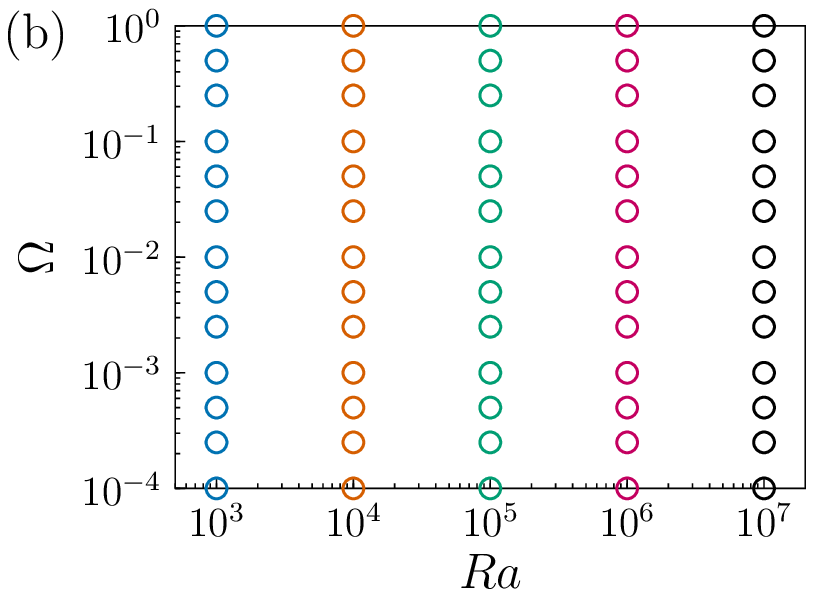}
\end{minipage}
\caption{(a) Sketch of the cylindrical domain and imposed TW. (b) Studied parameter space. The mesh sizes $n_r \times n_\varphi \times n_z$ of the DNS are $48 \times 130 \times 98$ for $Ra=10^3$, $96 \times 260 \times 196$ for $Ra=10^4,10^5,10^6$ and $128 \times 342 \times 256$ for $Ra=10^7$.}
\label{fig:rbc_setup}
\end{figure}

\subsection{Numerical Setup: Cylindrical Rayleigh-B{\'e}nard Convection (Pr=1)}
The setup is essentially motivated by the original experiments of \cite{Fultz1959}, where a heat source rotated around a cylinder with the radius $R$ (diameter $D$), except in our case thermal waves travel at the bottom and top and a mean temperature gradient was applied, as in Setup B of the previous part. In particular, the temperature distribution is linear in the radial $r$--direction and consists of one wave period in $\varphi$ that travels counterclockwise:
\begin{align*}
\theta(\varphi,r,z=0,t) &= 0.5\left[\frac{r}{R} \cos(\varphi - 2\pi \Omega t) + 1\right],\\  
\theta(\varphi,r,z=H,t) &= 0.5\left[\frac{r}{R} \cos(\varphi - 2\pi \Omega t) - 1\right].
\end{align*}
Again, the mean temperature gradient, averaged over time, is the same as in classical RBC. The cell is shown in figure \ref{fig:rbc_setup} (a). Furthermore, top and bottom plates are free-slip ($\partial \mathbf{u}/ \partial\mathbf{n} = 0$) and no-slip conditions are applied at the sidewall ($\mathbf{u}=0$). All simulations are carried out for the parameters $Pr=1$ and the aspect ratio $\Gamma\equiv D/H=1$. The rather large aspect ratio is a sacrifice, in return more simulations could be conducted and the parameter space in the region of interest is well resolved, as is shown in figure \ref{fig:rbc_setup} (b).

\subsection{Results}
Previously, we have shown that travelling thermal waves generate a mean horizontal, or, synonymous, a zonal flow. The same can be observed in the cylindrical system, where a zonal flow now refers to non-vanishing azimuthal mean flow. In the following, we evaluate its strength and direction and discuss the results in the context of the 2D results. As no specific adjustments to the theoretical model have been done, from that point on, the model results are intended to serve mainly as references to the previous results. A brief remark beforehand:  Evaluating the time and volume average of $u_\varphi$ proves problematic, as often flows are not purely pro- or retrograde.  Therefore, rather than to give precise scaling laws, the primary purpose of the subsequent analysis is to explore the parameter space, demonstrate the overall strength of the zonal flows, find the most critical wave frequencies and determine the critical $Ra$ above which the results deviate substantively from the predictions.\\

\begin{figure}
\begin{minipage}[c]{0.49\textwidth}
\centering
\includegraphics[width=0.99\textwidth]{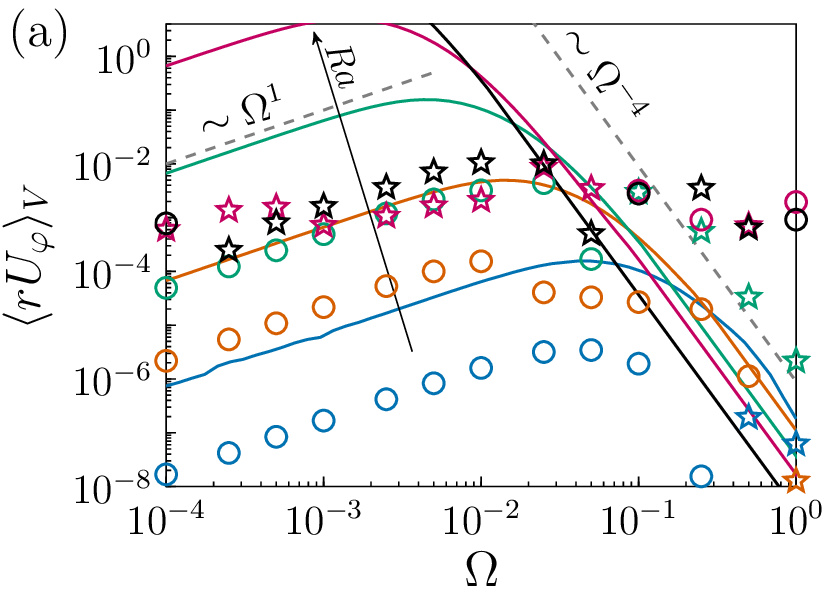}
\end{minipage}
\hfill
\begin{minipage}[c]{0.49\textwidth}
\centering
\includegraphics[width=0.99\textwidth]{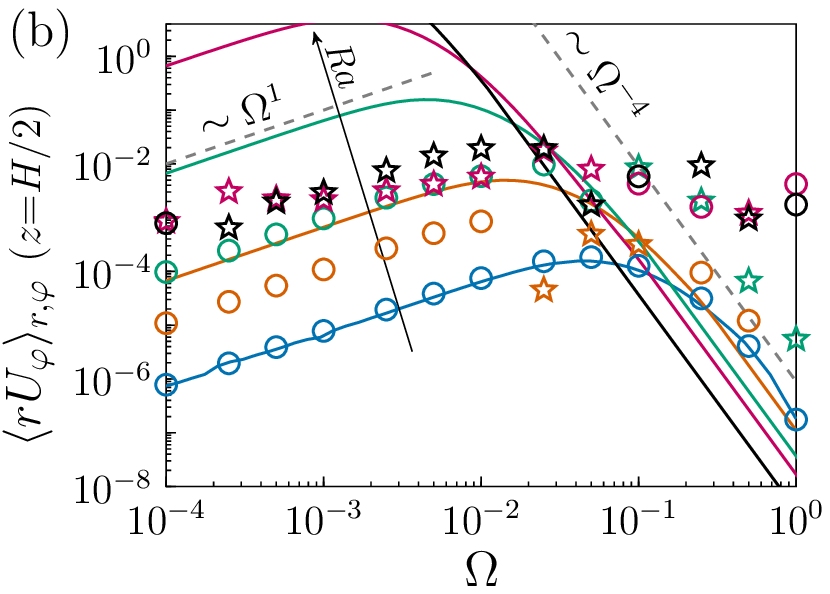}
\end{minipage}%
\caption{(a) Time- and volume averaged zonal flow as a function of the heat source frequency $\Omega$, (b) zonal flow at mid-height for 3D RBC data. $Ra =$ $10^3$ (blue), $10^4$ (orange), $10^5$ (green), $10^6$ (red) and $10^7$ (black). Circles (stars) denote a retrograde (prograde) mean zonal flow, the solid lines of the corresponding colour show the results of the theoretical model by \cite{Davey1967}. }
\label{fig:rbc_UvsOm}
\end{figure}
Figure \ref{fig:rbc_UvsOm} shows (a) the total mean azimuthal momentum $\langle U_\varphi \rangle_V$ and (b) the value of $\langle U_\varphi \rangle_{r,\varphi}$ at the mid-height. As before, circles denote a retrograde, stars a prograde mean flow, and the solid lines are the 2D model solutions from \cite{Davey1967}, without modifications for no-slip walls. The obtained flows for small $Ra\leq 10^5$ share distinct features with the 2D flows. The mean momentum converges to the asymptotic scalings, and, in fact, the data of figure \ref{fig:rbc_UvsOm} b collapse under a transformation with $Ra$ remarkably good. For larger $\Omega$, in particular $\Omega \geq 10^{-1}$, the most flows are found to be directed prograde, even for $Ra=10^3$, which is different from the 2D case. And as in 2D, the flow structures reveal a transition in this $\Omega$--region. As was discussed in section \ref{sec:spacetime}, the plane of the LSC drifts with the same speed as the TW ($=\Omega$), if the TW speed is small compared to thermal diffusion speed, and the LSC breaks off from the TW at larger $\Omega$, forming separate structures, acting on different timescales. It is in the regime of this break-off above which a prograde flow is present. This process hints towards a similar mean flow instability, as discussed in §\ref{sec:2d-prograde}, where the mean flow is now a slow LSC. 

\begin{figure}
\begin{minipage}[c]{0.195\textwidth}
\centering
\includegraphics[width=0.99\textwidth,trim=5 0 5 0,clip]{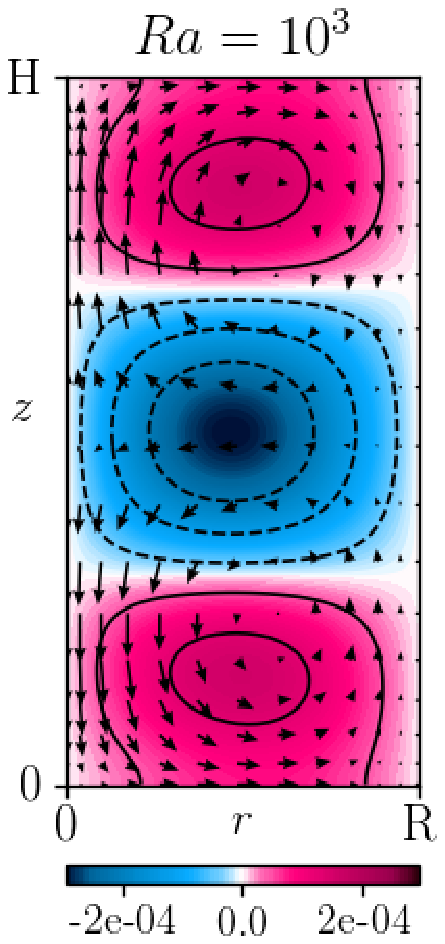}
\end{minipage}
\begin{minipage}[c]{0.195\textwidth}\centering
\includegraphics[width=0.99\textwidth,trim=5 0 5 0,clip]{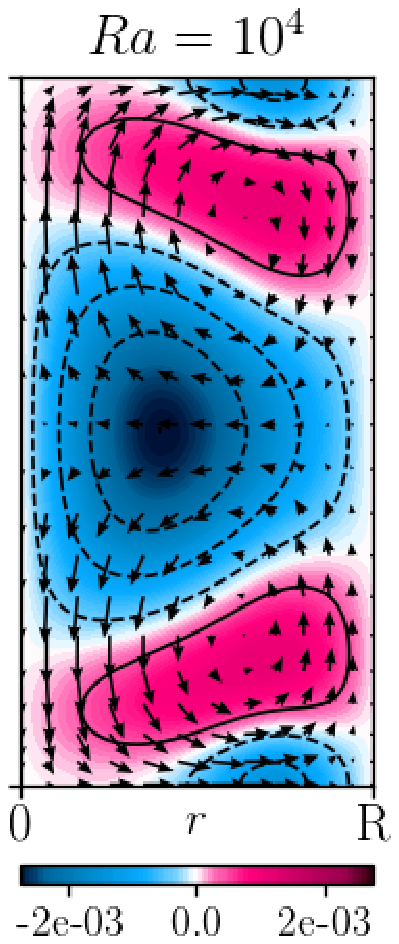}
\end{minipage}
\begin{minipage}[c]{0.195\textwidth}\centering
\includegraphics[width=0.99\textwidth,trim=5 0 5 0,clip]{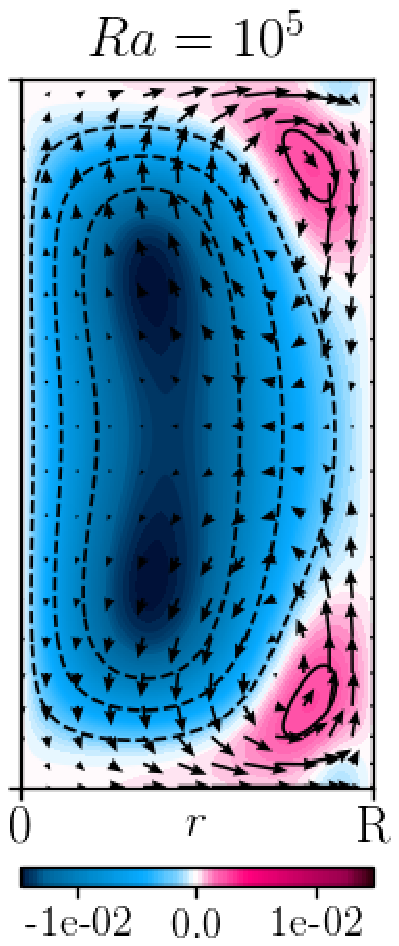}
\end{minipage}
\begin{minipage}[c]{0.195\textwidth}\centering
\includegraphics[width=0.99\textwidth,trim=5 0 5 0,clip]{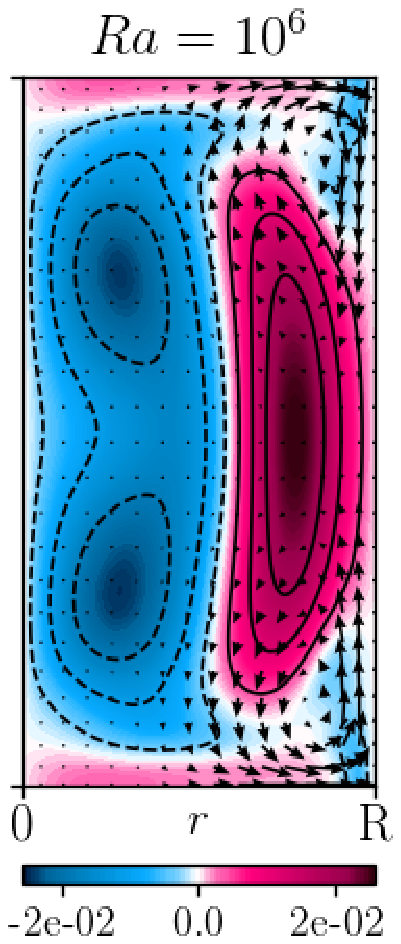}
\end{minipage}
\begin{minipage}[c]{0.195\textwidth}\centering
\includegraphics[width=0.99\textwidth,trim=5 0 5 0,clip]{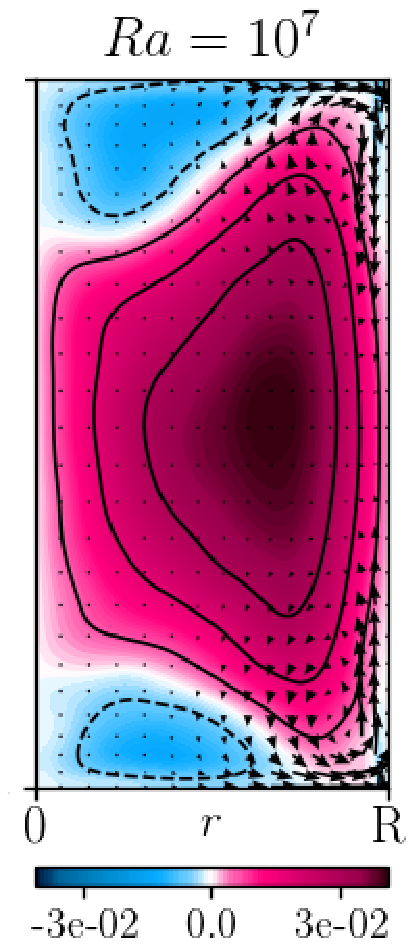}
\end{minipage}
\begin{minipage}[c]{0.195\textwidth}\centering
\includegraphics[width=0.99\textwidth,trim=50 0 50 0,clip]{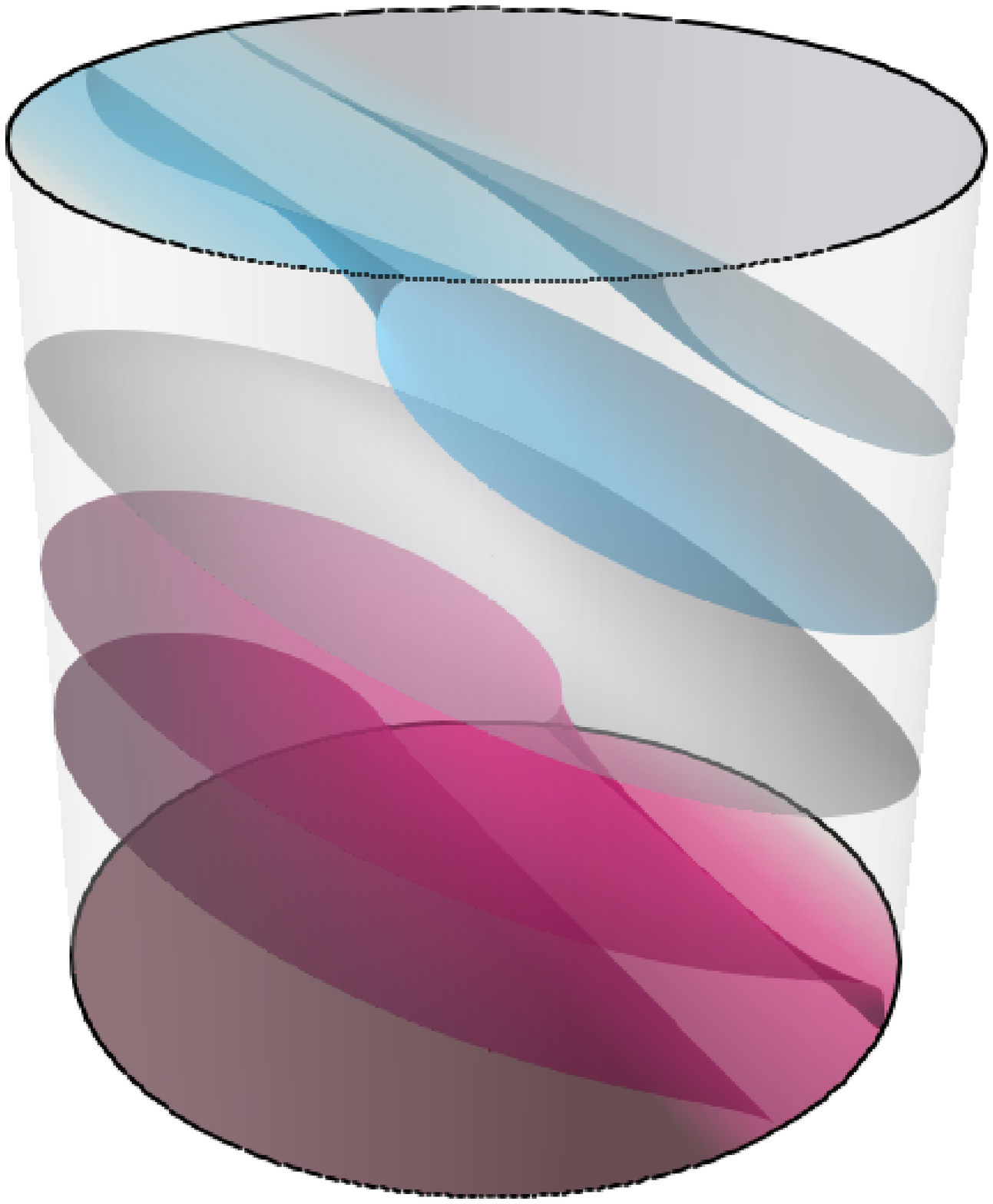}
\end{minipage}
\begin{minipage}[c]{0.195\textwidth}\centering
\includegraphics[width=0.99\textwidth,trim=50 0 50 0,clip]{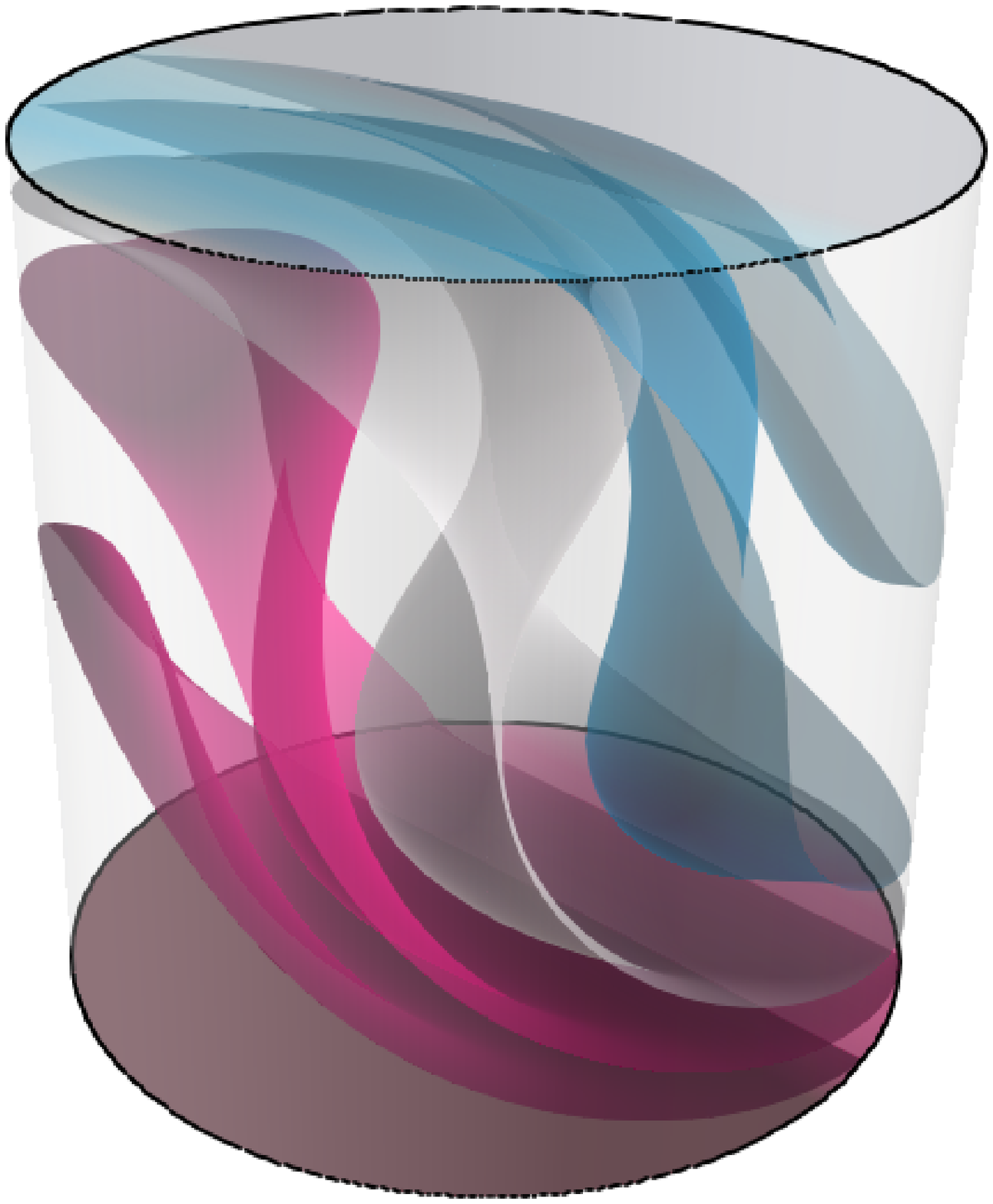}
\end{minipage}
\begin{minipage}[c]{0.195\textwidth}\centering
\includegraphics[width=0.99\textwidth,trim=50 0 50 0,clip]{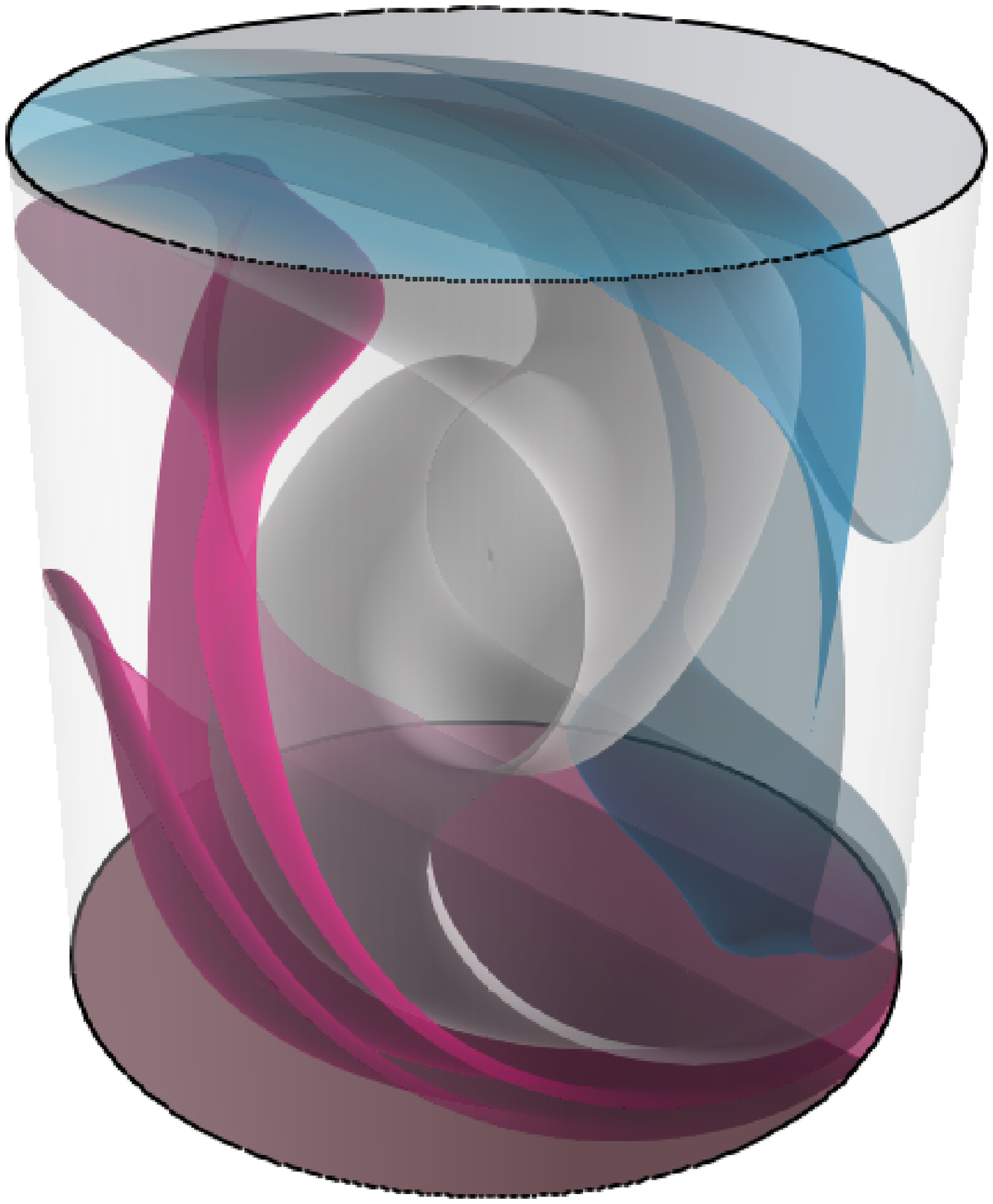}
\end{minipage}
\begin{minipage}[c]{0.195\textwidth}\centering
\includegraphics[width=0.99\textwidth,trim=50 0 50 0,clip]{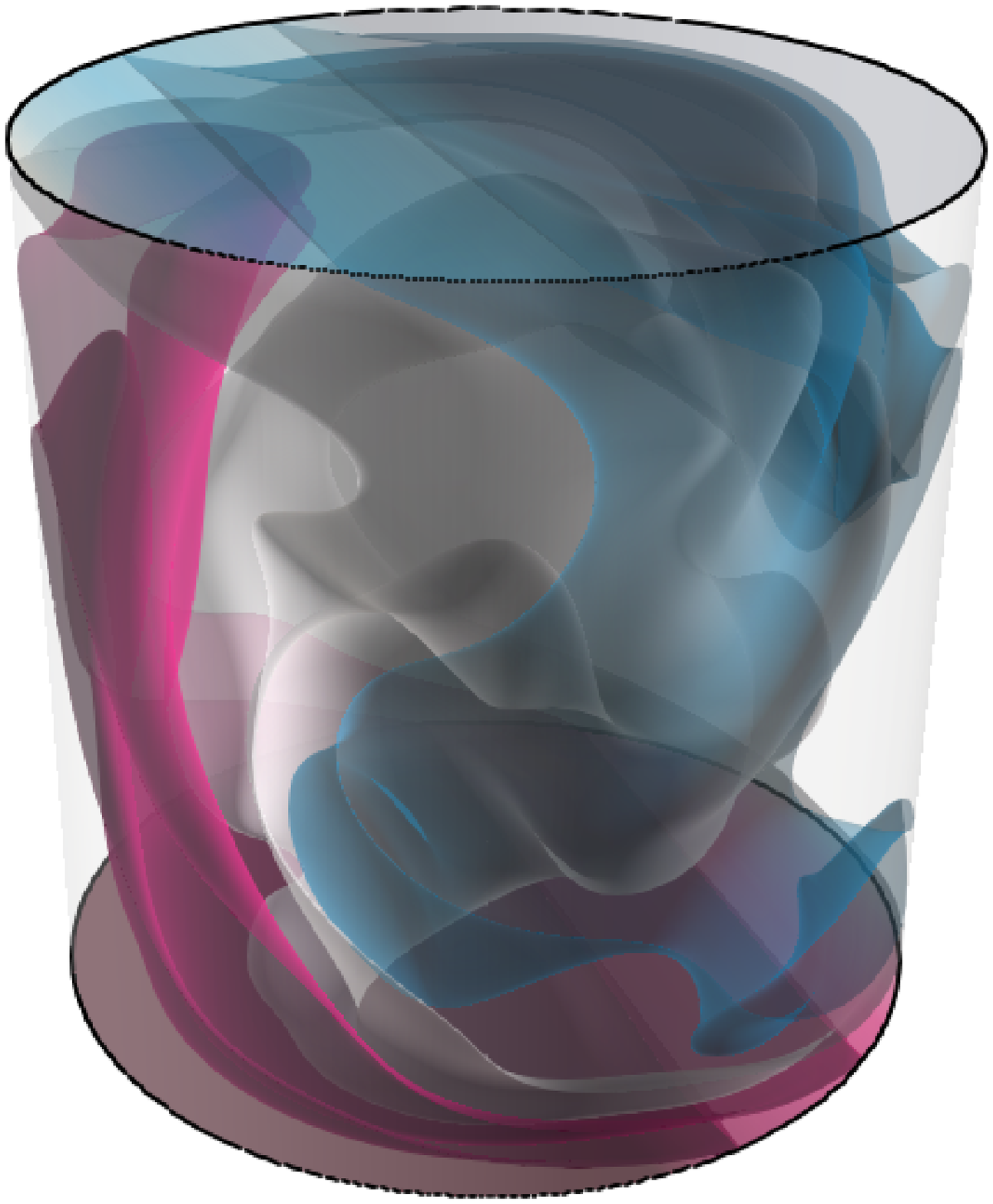}
\end{minipage}
\begin{minipage}[c]{0.195\textwidth} \centering
\includegraphics[width=0.99\textwidth,trim=50 0 50 0,clip]{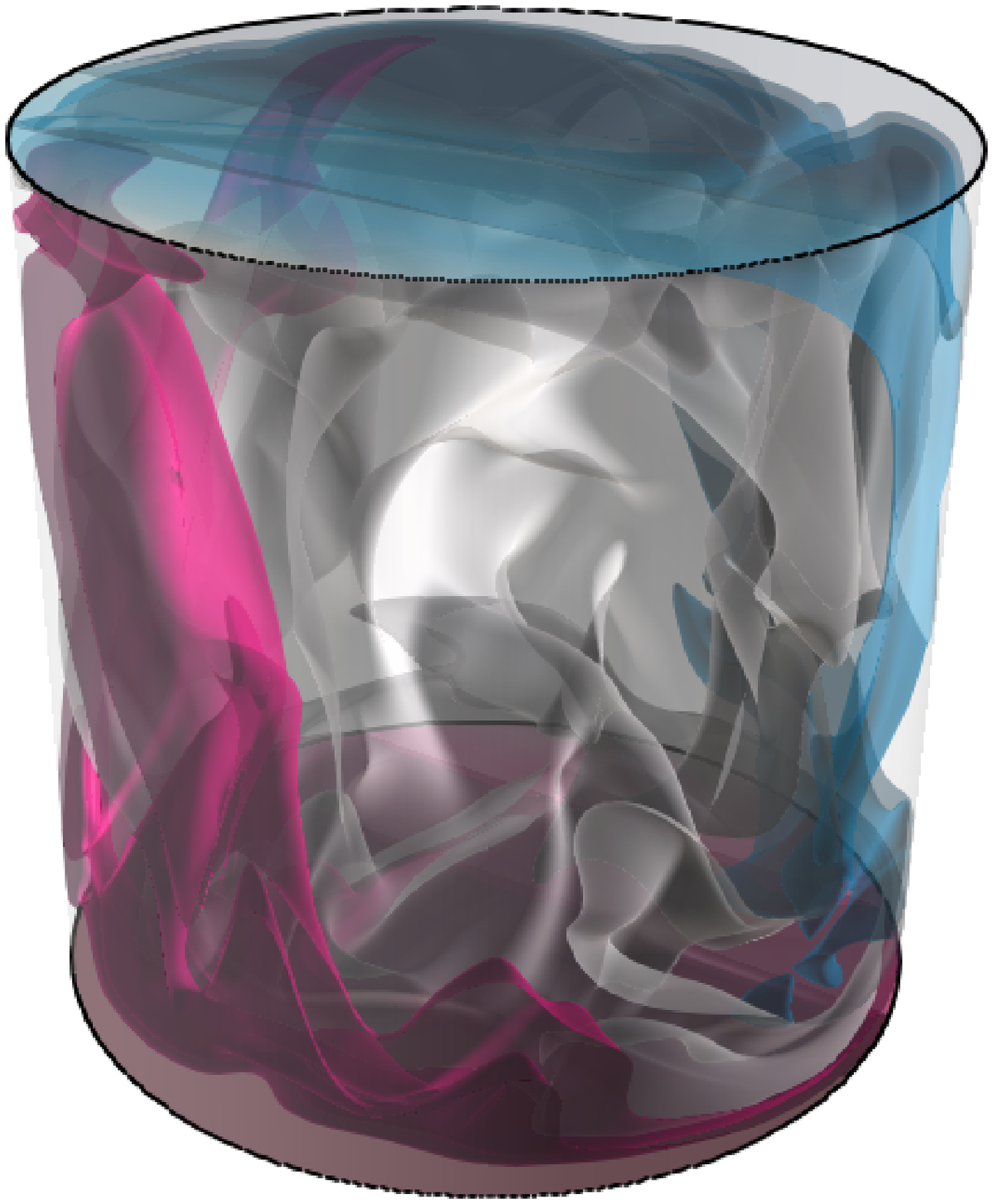}
\end{minipage}
\caption{ For a fixed TW frequency $\Omega$ $=$ $0.01$.  The azimuthally averaged mean azimuthal velocity $\langle U_\varphi \rangle_\varphi$ (top row) and the corresponding snapshots of the temperature $\theta$ (bottom row). As $Ra$ increases, the core zonal flow becomes first stronger retrograde ($Ra=10^4,10^5$), then switches its state to a prograde flow originated from the sidewall ($Ra \geq 10^6$), while still increasing its strength (see colour bar).}
\label{fig:rbc_fields}
\end{figure}

As $Ra$ exceeds $10^5$, turbulent fluctuations increase and the data in figure \ref{fig:rbc_UvsOm} becomes increasingly scattered. The asymptotic scalings are hardly determinable, even though $\langle U_x\rangle_V\sim \Omega^1$ for $\Omega \rightarrow 0$ appear still valid. The fluctuations can exceed their mean values, especially for small and large $\Omega$. Despite the strong fluctuations, in regions of maximal zonal flow, i.e. $\Omega \approx 10^{-2}$, the mean values are highly significant and can induce zonal flows of the same order of magnitude as the TW frequency, $\langle U_\varphi \rangle_V \approx \mathcal{O}( 10^{-2})$. Furthermore, similarly to the 2D case, in 3D, the zonal flows at high $Ra$ are most of the time directed prograde, contrary to small $Ra$. From the vertical planes of the azimuthally and time averaged azimuthal velocity, shown in figure \ref{fig:rbc_fields}, the dominance of prograde motion at large $Ra$ becomes more obvious. Moreover, these figures reveal a complex, inhomogeneous flow, with strong differential rotation and poloidal mean velocities. 

\subsubsection{Vertical and radial momentum transport}
In the following, we assess the contributing terms of the mean flow azimuthal momentum equation. For clarity, let us write the equation for $u_\varphi$ explicitly:
\begin{align}
\begin{split}
   \partial_t u_\varphi + \frac{1}{r} \frac{\partial r u_\varphi u_r}{\partial r} + & \frac{1}{r} \frac{\partial u_\varphi u_\varphi}{\partial \varphi} + \frac{\partial u_\varphi u_z}{\partial z} =  \\ 
\frac{\partial p}{\partial \varphi} & + \sqrt{\frac{Pr}{Ra}} \left[ \frac{1}{r} \frac{\partial}{\partial r}\left( r \frac{\partial u_\varphi}{\partial r}\right)  + \frac{1}{r^2}\frac{\partial^2 u_\varphi}{\partial \varphi^2} +\frac{\partial^2 u_\varphi}{\partial z^2} - \frac{u_\varphi}{r^2} + \frac{2}{r^2}\frac{\partial u_r}{\partial \varphi}\right].
\end{split}
\label{eq:uphi}
\end{align}

First, we consider how $U_\varphi$ changes in the vertical direction and, second, how it changes radially. Therefore, decomposing eq. (\ref{eq:uphi}) into its mean and fluctuation components, and averaging over $\varphi$ and $r$ gives the following balance:
\begin{align}
\begin{split}
 \sqrt{\frac{Pr}{Ra}}  \left( {\frac{\partial^2 \langle U_\varphi \rangle_{r,\varphi}}{\partial_z^2}} -\frac{\langle U_\varphi \rangle_{r,\varphi}}{r^2} \right) & = \\
  \frac{\partial \langle \overline{u_\varphi^\prime u_z^\prime} \rangle_{r,\varphi}}{\partial_z} & + 
 \frac{\partial \langle   U_\varphi  U_z \rangle_{r,\varphi}}{\partial_z}
 + \langle \frac{ \overline{u_\varphi^\prime u_r^\prime }}{r}\rangle_{r,\varphi} +  \langle \frac{U_r U_\varphi }{r}\rangle_{r,\varphi}.
\end{split}
\label{eq:rbc_balance_z}
\end{align}

Analysing the radial dependence, on the other side, averaging over $\varphi$ and $z$ gives  
\begin{align}
\begin{split}
 \sqrt{\frac{Pr}{Ra}} \left(  \frac{1}{r^2} {\frac{\partial^2 \langle U_\varphi \rangle_{\varphi,z}}{\partial_r^2}}  -\frac{\langle U_\varphi \rangle_{\varphi,z}}{r^2} \right) & = \\
  \frac{1}{r}\frac{\partial r\langle \overline{u_\varphi^\prime u_r^\prime} \rangle_{\varphi,z}}{\partial_r} & + 
  \frac{1}{r}\frac{\partial r\langle   U_\varphi  U_r \rangle_{\varphi,z}}{\partial_r}
 + \langle \frac{ \overline{u_\varphi^\prime u_r^\prime }}{r}\rangle_{\varphi,z} +  \langle \frac{U_r U_\varphi }{r}\rangle_{\varphi,z}.
\end{split}
\label{eq:rbc_balance_r}
\end{align}

\begin{figure}
\begin{minipage}[]{0.98\textwidth}
\centering
\includegraphics[width=0.99\textwidth,trim=0 0 0 0,clip]{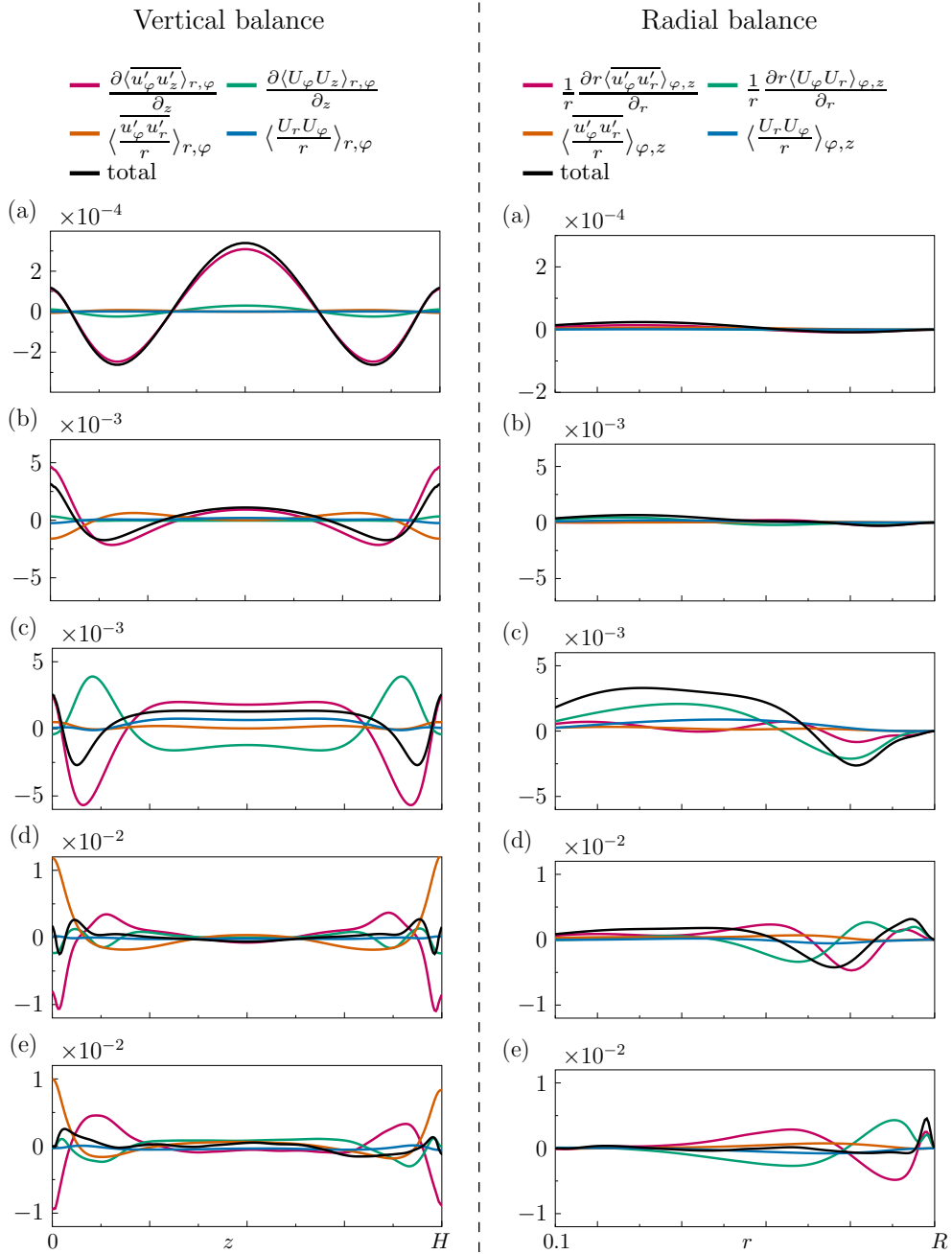}
\end{minipage}
\caption{Components of the vertical momentum transport, eq. (\ref{eq:rbc_balance_z}), (left) and the radial momentum transport, eq. (\ref{eq:rbc_balance_r}), (right). Parameters: $\Omega=10^{-2}$ and $Ra$: (a) $10^3$, (b) $10^4$, (c) $10^5$, (d) $10^6$ and (e) $10^7$.}
\label{fig:rbc_balance}
\end{figure}

The rhs-terms of these equations are evaluated for $\Omega=10^{-2}$, which are shown in figure \ref{fig:rbc_balance}. We ensured that in the simulations the data was averaged over an integer number of the TW periods, to prevent artifacts of the TW in the mean fields (the exact time values can be found in the supplementary material). When we compare the individual mean velocities for (a) $Ra=10^3$ and (b) $Ra=10^4$, it becomes clear that the mean field transport in both, vertical and radial, directions is rather negligible. Hence, the non-linear Reynolds stress sustains the mean zonal flow, just like in the 2D case for small $Ra$ (see figure \ref{fig:2d_balance} a), and as expected \citep{Stern1959,Davey1967}. The small mean field contributions even reinforce the zonal flow, since the shape of the mean advection curves matches the shape of the Reynolds stress curve. Comparing further the vertical and the radial transport, we find that the former dominates the latter one by an order of magnitude. This proves that in this case the neglect of the radial currents, as suggested by \cite{Stern1959}, is justified, and therefore the mean momentum scalings (figure \ref{fig:rbc_UvsOm}) match remarkably well with its 2D analogue (figure \ref{fig:2d_UvsOm}), and the difference in the prefactors can presumably be explained by the different velocity BCs.

The situation for larger $Ra$ (figure \ref{fig:rbc_balance} c-e) is vastly different. First, the problem becomes considerably three dimensional and the radial transport now reaches the same order of magnitude as the vertical transport (e.g. figure \ref{fig:rbc_balance} c-e), which suggests that the validity of the 2D analogy at large $Ra$ is no longer justified. Furthermore, the mean field advection contributions, which can be partially seen from figure \ref{fig:rbc_fields}, increase significantly. As a matter of fact, locally it can even exceed the Reynolds stress contributions. Furthermore, whereas for small $Ra$, vertical and radial momentum transports are present throughout the whole domain, at large $Ra$ it becomes strongly confined to the boundaries. Especially the vertical transport peaks close to the top and bottom boundaries and is less pronounced in the center. The radial transport, on the other side, shows an interesting feature in the region $0.95 \leq r/R \leq 1$ (figure \ref{fig:rbc_balance} d,e). All terms are simultaneously positive, which causes an enhanced zonal transport close to the sidewall. This may explain why a prograde flow first appears close to the sidewall (figure \ref{fig:rbc_fields}, $Ra = 10^6$) and from there spreads further inwards (figure \ref{fig:rbc_fields}, $Ra = 10^7$).

\subsubsection{Sensitivity on the boundary conditions and aspect ratio}
\begin{figure}
\begin{minipage}[c]{0.995\textwidth}
\centering
\includegraphics[width=0.55\textwidth,trim=0 0 0 0,clip]{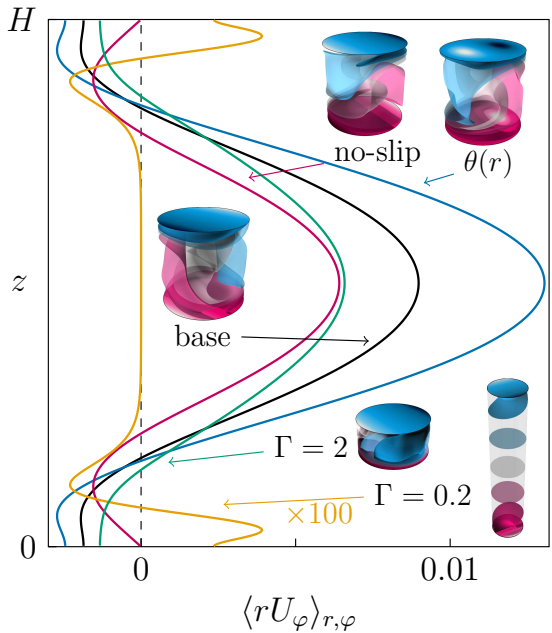}
\end{minipage}
\caption{ Mean angular momentum profile for $Ra=10^5$ and $\Omega=10^{-1}$. The curves show the effects of different imposed BCs and aspect ratios: Baseline simulation (black, free-slip BCs, $\theta\sim r/R$ and $\Gamma=1$), sinusoidal radial temperature BCs (blue, free-slip BCs, $\theta\sim sin(\pi r/R)$ and $\Gamma=1$), no-slip (red, no-slip BCs, $\theta\sim r/R$ and $\Gamma=1$), $\Gamma=0.2$ (yellow, free-slip BCs, $\theta\sim r/R$ and $\Gamma=0.2$) and $\Gamma=2$ (green, free-slip BCs, $\theta\sim r/R$  and $\Gamma=2$).}
\label{fig:configurations}
\end{figure}

 The systems studied in this paper allow many variations of the velocity and temperature boundary conditions as well as geometrical characteristics of the system. Discussing all of them goes beyond the scope of a single study. Nevertheless, in order to provide some preliminary intuition, we examine selected variations and their effects on the generation of the zonal flows. We do this for a single baseline simulation at $Ra=10^5$ and $\Omega=10^{-1}$. The mean angular momentum profiles are shown in figure \ref{fig:configurations}.

First we consider the effects of the aspect ratio. From classical RBC it is known that zonal flow properties depend strongly on $\Gamma$ \citep{Wang2020a}. In our case, a decrease of the aspect ratio from $\Gamma=1$ to $\Gamma=0.2$ (slender cell) weakens the zonal flow considerably by a factor of $100$. Furthermore, the zonal flow becomes confined to the top and bottom plates, while no zonal flow is observed in the center of the cell. On the other hand, increasing the aspect ratio to $\Gamma=2$ has only minor impact on the zonal flow. We must note that for the case of $\Gamma=0.2$, convection has yet not started and subsequent studies would be necessary to conclusively elucidate on the aspect ratio dependence.

The effects of the boundary condition variations on the formation of zonal flows can be formulated as follows. No-slip conditions at the top and bottom plates lead to a slightly weaker, but qualitatively similar zonal flow. Likewise, replacing the linear radial temperature distribution at the plates by a sinusoidal distribution ($\theta\sim sin(\pi r/R)$) shows still a qualitatively similar angular momentum profile, though the strength of the zonal flow in the center of the cell increases by a factor of about $1.5$. This indicates that the system is rather sensitive to variations of the temperature BCs.

\subsection{Example: Atmospheric boundary layer}

Finally, we would like to illustrate the strength of the induced zonal flows on a concrete example. Assume an atmospheric boundary layer with a height of $\hat{H}=500m$ and a vertical temperature difference of $\Delta T=3^\circ $C. Given a mean temperature of $10^\circ $C, the material properties of air are approximately $\kappa=2.0 \times 10^{-5}$ m$^2/$s, $\nu=1.4 \times 10^{-5}$ m$^2/$s and $\alpha=3.6 \times 10^{-3}$K$^{-1}$. From that we find $Pr\approx 0.7$ and $Ra\approx10^{16}$ and the free-fall units $u_{ff} \equiv \sqrt{\alpha g \hat{H} \Delta \theta}\approx 7$m$/$s, $t_{ff} \equiv \hat{H}/u_{ff} \approx 70$s. This system is exposed to a travelling thermal wave through the solar radiation with a period of $24$h, or, in dimensionless units $\Omega \approx 10^{-3}$. For simplicity, we say, the day and night difference is also about $3^\circ $C, which is likely to be a rather conservative estimate. Our study does not conclusively show, how the zonal flows scale up to $Ra=10^{16}$, but the results suggest a saturation at higher $Ra$, therefore we proceed using the maximum order of magnitude, which is $U_\varphi \approx 10^{-2}$ (for the given $\Omega$ it might be smaller). With these values, the thermal variation of the Earth's surface would induce a prevailing zonal flow of around $0.07$m$/$s, or equivalently $0.3$km$/$h. However, locally it could exceed this value (see figure \ref{fig:rbc_fields}) multiple times, therefore speeds of $1$km$/$h are conceivable. Nevertheless, the variance of this estimate is rather high. Subsequent studies have to examine the influence of $Ra,Pr$ and the geometry, in order to make more confident statements about natural systems.

\section{Conclusions}
\label{sec:conclusion}

We have explored the original moving heat source problem by means of direct numerical simulations in 2D and 3D systems, for varying Rayleigh numbers $Ra$ and travelling thermal wave (TW) frequency $\Omega$. In the seminal works of \cite{Fultz1959} and \cite{Stern1959}, it was discovered that a system subjected to such a travelling wave generates Reynolds stresses, which induce a large scale mean horizontal, or equivalently zonal, flow directed counter to the propagating thermal wave. Therefore, in the first part, we revisited the theoretical model proposed by \cite{Davey1967} and found excellent agreement with the theory for low $Ra$ flows, where even the absolute magnitude of the zonal flows is reproduced remarkably good. As $Ra$ increases, the theoretical model overestimates the DNS data, which is consistent with the effects of higher order non-linear contributions \citep{Whitehead1972,Young1972,Hinch1971}. 

However, when an unstable mean temperature gradient is added to the system, the flows deviate substantially from the initial predictions and often reverse their direction to a prograde moving zonal flow. Such a behaviour was theorised before, as the result of a mean flow instability caused by the tilt of convection cells \citep{Thompson1970,Busse1972, Busse1983}. Therefore, we have conducted a global linear stability analysis of a base flow near onset of convection and confirmed this hypothesis. The most unstable mode can give rise to a reverse of the horizontal velocity profile. Despite the strong plausibility, that this mean flow instability is the dominating mechanism at large $Ra$, the question remains open why prograde flow are more numerous than retrograde flows, while the mean flow instability suggests a spontaneous break of symmetry and therefore a more balanced distribution. In this context, it would be interesting to study  in the future the interaction between the TW induced and convection rolls induced fields.

In the second part we have examined the moving heat source problem in the context of a 3D cylindrical RBC system. The asymptotic scalings $\langle U_\varphi \rangle_V\sim \Omega^1$ for $\Omega \rightarrow 0$ and $\langle U_\varphi \rangle_V\sim \Omega^{-4}$ for $\Omega \rightarrow \infty$ of the 2D theoretical model \citep{Davey1967} still hold in this system, especially at small $Ra$. An analysis of the vertical and radial momentum transport contributions suggests that the radial transport is negligible at small $Ra$, (which justifies a 2D approximation) but becomes relevant as $Ra$ increases. Furthermore, again, large $Ra$ are found predominantly inducing a prograde mean zonal flow. This gives more evidence that the prograde prevalence is likely not fully explained by the mean flow instability picture and further study's are required to explain its origin. 

The studied problem is sufficiently general enough and can be extended to more complicated systems \citep{Whitehead1975,Shukla1981,Mamou1996}. A more generalized theoretical framework already exist, which includes the influence of a basic stability and rotation \citep{Stern1971,Chawla1983}, however, as this study showed, the theoretical models most often cannot fully explain the phenomena in convection dominated systems. Furthermore, the moving heat source problem might help to understand ubiquitous structures present in rotating systems. In rotating RBC systems, the flow structures near the sidewall \citep{Favier2020,Zhang2020} are similar to a certain extent to those structures accounting from the imposed TW.

Ultimately, this study also revealed that the estimate of the order of magnitudes is still afflicted with too large variances to make reliable statements about natural systems. A naive approach showed that atmospheric currents, caused by solar radiation and Earth's rotation, can actually generate prevailing zonal flows of about $1.0$ km$/$h. However, the variance of this estimate is rather high, it therefore is pivotal for subsequent studies to examine the sensitivities with $Ra,Pr$ and geometry in greater detail.

\section*{Acknowledgements}
We acknowledge the support by the Deutsche Forschungsgemeinschaft (DFG) under the grant Sh405/10 and Sh405/7 (SPP 1881 “Turbulent Superstructures”) and the Leibniz Supercomputing Centre (LRZ). We also thank the Max-Planck HPC Teams in G{\"o}ttingen and Munich for their generous technical support and additional computational resources.
\section*{Declaration of interests.}
The  authors  report  no  conflict  of  interest.



\begin{appendix}
\section{Movies}
\textit{Movie 1:}
Contours of the temperature field for $Ra=10^5$ of Setup B. The arrow below each simulation indicates the speed of the travelling wave. Travelling wave speeds from top to bottom: $\Omega=0.01, 0.1$ and $1.0$.\\

\textit{Movie 2:}
Contours of the temperature field for $Ra=10^6$ of Setup B. The arrow below each simulation indicates the speed of the travelling wave. Travelling wave speeds from top to bottom: $\Omega=0.01, 0.1$ and $1.0$.\\

\textit{Movie 3:}
Contours of the temperature field for $Ra=10^7$ of Setup B. The arrow below each simulation indicates the speed of the travelling wave. Travelling wave speeds from top to bottom: $\Omega=0.01, 0.1$ and $1.0$.\\

\textit{Movie 4:}
Contours of the temperature field for $Ra=10^5$ of Setup B. The arrow below each simulation indicates the speed of the travelling wave. Prograde moving plumes are observed for $\Omega=0.1$ (top) and retrograde moving plumes for $\Omega=0.316$ (bottom).\\

\textit{Movie 5:}
Contours of the temperature field and time evolution of the zonal flow $\langle u_x \rangle $ for $Ra=10^6$ and $\Omega=0.1$. The arrow below each simulation indicates the speed of the travelling wave. Retrograde moving structures are observed for Setup A (top) and prograde moving structures for Setup B (bottom).\\

\textit{Movie 6:}
Contours of the temperature field for $Ra=10^5$ of the 3D cylindrical system. Travelling wave speeds from left to right: $\Omega=0.01, 0.1$ and $1.0$.\\

\textit{Movie 7:}
Contours of the temperature field for $Ra=10^6$ of the 3D cylindrical system. Travelling wave speeds from left to right: $\Omega=0.01, 0.1$ and $1.0$.\\

\textit{Movie 8:}
Contours of the temperature field for $Ra=10^7$ of the 3D cylindrical system. Travelling wave speeds from left to right: $\Omega=0.01, 0.1$ and $1.0$.

\section{Theory for diffusion dominated flows}
\label{app:theory}

We follow the theory of \cite{Davey1967}, but solve the equations in a more general way, to allow for flexibility in the chosen BCs; for more details, the reader is referred to \cite{Davey1967} or \cite{Kelly1970}.
Neglecting the mean vertical velocity component, assuming the mean horizontal velocity to be independent of $x$ and neglecting the contributions from the mean temperature field $\overline{\theta}$, the linearized, non-dimensionalised Navier-Stokes equations in two-dimensions read
\begin{align}
\partial_t u^\prime + (U+u^\prime)\partial_x u^\prime + w^\prime \partial_z (U+u^\prime) &= -\partial_x p + \nu^{*}
\left( \frac{\partial^2 U}{\partial z^2} + \frac{\partial^2 u^\prime}{\partial x^2} + \frac{\partial^2 u^\prime}{\partial z^2}\right) \label{eq:appu},\\
\partial_t w^\prime + (U+u^\prime)\partial_x w^\prime + w^\prime\partial_z (w^\prime) &= -\partial_z p + \nu^* 
\left(\frac{\partial^2 w^\prime}{\partial x^2} + \frac{\partial^2 w^\prime}{\partial z^2}\right) + \theta^\prime \label{eq:appw}, \\
\partial_x u^\prime + \partial_z w^\prime = 0.
\end{align}

Here, $u^\prime$ and $w^\prime$ are, respectively, horizontal and vertical components of the velocity fluctuations with respect to its time-average, i.e. $U$ and $W=0$, and $\theta^\prime$ is the temperature fluctuation. For non-dimensionalisation we have used the free-fall velocity  $u_{ff}\equiv (\alpha g \Delta \hat{H})^{1/2}$, the height $\hat{H}$ and the amplitude of the thermal TW, $\Delta$, so that $\nu^*=\sqrt{Pr/Ra}$. Let us consider a single wave mode in the horizontal $x$-direction and in time $t$, e.g.:
 \begin{align}
w^\prime(x,z,t) &= \frac{1}{2}\left( \hat{w}(z)e^{+i(kx-2\pi \Omega t)} +\hat{w}^*(z)e^{-i(kx-2\pi \Omega t)} \right),\label{eq:sw_w}\\
u^\prime(x,z,t) = -\int \partial_z w^\prime dx &= \frac{i}{2k}\left( \partial_z\hat{w}(z)e^{+i(kx-2\pi \Omega t)} -\partial_z\hat{w}^*(z)e^{-i(kx-2\pi \Omega t)} \right), \label{eq:sw_u}\\
\theta^\prime(x,z,t) &= \frac{1}{2}\left( \hat{\theta}(z)e^{+i(kx-2\pi \Omega t)} +\hat{\theta}^*(z)e^{-i(kx-2\pi \Omega t)} \right), \label{eq:sw_temp}
 \end{align}
where the asterisk denotes the complex conjugate of a function. 
We will consider two BCs (different scenarios), \textit{Scenario 1} describes a setup, where two travelling thermal waves are imposed at the top and the bottom (whithout any phase difference). This case was considered in the present work. \textit{Scenario 2}, on the other hand, describes a setup, where the thermal wave travels only at the bottom, while the dimensionless top temperature equals zero.\\

\textbf{Step 1}: Calculate $\hat{\theta}(z)$.

Neglecting dissipation in $x$, all convective terms and mean temperature contributions, the linearized non-dimensional energy equation reads
\begin{align*}
\partial_t \theta^\prime  &= \kappa^*\left(\frac{\partial^2 \theta^\prime}{\partial z^2}\right),
\end{align*}
where  $\kappa^*=1/\sqrt{RaPr}$. This, together with eq. (\ref{eq:sw_temp}), leads to the following equation for the wave amplitude equation $\hat{\theta}(z)$:
\begin{align}
\frac{d^2 \hat{\theta}}{d z^2} - \lambda^2 \hat{\theta} = 0; \quad  \lambda^2 = \frac{2\pi i\Omega}{\kappa^*}.
\label{eq:hatT}
\end{align}
The solution to eq. (\ref{eq:hatT}), for the two scenario's, is:

\begin{minipage}[t]{0.48\textwidth}
\centering
\textit{Scenario 1}
\begin{align*}
&\text{For}\ \hat{\theta}\vert_{z=-1/2} = \hat{\theta}\vert_{z=1/2} = \frac{1}{2} :\\
&\hat{\theta}(z) =  \frac{\cosh(\lambda z)}{2 \cosh(\lambda/2)} .
\end{align*}
\end{minipage}
\begin{minipage}[t]{0.48\textwidth}
\centering
\textit{Scenario 2}
\begin{align*}
&\text{For}\ \hat{\theta}\vert_{z=-1/2} = \frac{1}{2}, \hat{\theta}\vert_{z=1/2} = 0 :\\
&\hat{\theta}(z) =  \frac{\sinh(\lambda/2 - \lambda z)}{2 \sinh(\lambda)} .
\end{align*}
\end{minipage}\\

\textbf{Step 2}: Calculate $\hat{w}(z)$.

Eliminate the pressure term by cross differentiation of (\ref{eq:appu}) and (\ref{eq:appw}), substitute (\ref{eq:sw_w})-(\ref{eq:sw_temp}), neglect convective terms and assume that the thermal wavelength is much larger than the height of the cell ($kH\ll 1$) to obtain
\begin{align}
\frac{\partial^4 \hat{w}}{\partial z^4} - \alpha^2 \frac{\partial^2 \hat{w}}{\partial z^2} = k^2\hat{\theta}, \quad \alpha^2 = \frac{2\pi i\Omega}{\nu^*}.
\label{eq:hatw}
\end{align}
For $\hat{w}\vert_{z=1/2} = \hat{w}\vert_{z=-1/2} = \partial_z \hat{w}\vert_{z=1/2} = \partial_z \hat{w}\vert_{z=-1/2} = 0$, the solution to (\ref{eq:hatw}) is:

\begin{align*}
&\hat{w}(z) = \frac{c_1}{\alpha^2}\cosh(\alpha z) +\frac{c_2}{\alpha^2}\sinh(\alpha z) + c_3 z + c_4 + c_5 \cosh(\lambda z) + c_6 \sinh(\lambda z).
\end{align*}

{\allowdisplaybreaks
\begin{minipage}[t]{0.48\textwidth}
\centering
\textit{Scenario 1}
\begin{align*}
A &= \frac{k^2}{2\nu^* \lambda^2(\lambda^2-\alpha^2)}, \\
c_1 &= - \lambda\alpha A \frac{\tanh(\lambda/2)}{\sinh(\alpha/2)},  \\
c_2 &= 0, \\
c_3 &= 0, \\
c_4 &= A\left( \frac{\lambda}{\alpha} \frac{\tanh(\lambda/2)}{\tanh(\alpha/2)} -1\right), \\
c_5 &= \frac{A}{\cosh(\lambda/2)}, \\
c_6 &= 0.
\end{align*}
\hfill
\end{minipage}
\begin{minipage}[t]{0.48\textwidth}
\centering
\textit{Scenario 2}
\begin{align*}
A &= \frac{k^2}{4\nu^* \lambda^2(\lambda^2-\alpha^2)}, \\
c_1 &= - \lambda\alpha A \frac{\tanh(\lambda/2)}{\sinh(\alpha/2)},  \\
c_2 &= \frac{-\alpha A  \left(\frac{\lambda}{\tanh(\lambda/2)} -2\right)}{(2/\alpha) \sinh(\alpha/2) - \cosh(\alpha/2)}, \\
c_3 &=  -\frac{c_2}{\alpha}\cosh(\alpha/2) + \frac{\lambda A}{\tanh(\lambda/2)}, \\ 
c_4 &= A\left( \frac{\lambda}{\alpha} \frac{\tanh(\lambda/2)}{\tanh(\alpha/2)} -1\right), \\
c_5 &= \frac{A}{\cosh(\lambda/2)}, \\
c_6 &= \frac{-A}{\sinh(\lambda/2)}.
\end{align*}
\end{minipage}
}

\textbf{Step 3}: Calculate $U(z)$. 

Averaging equation (\ref{eq:appu}) over time and over one wavelength in $x$, we obtain the following equation for the mean flow $U(z)$:
\begin{align}
\label{eq:step3}
\nu^* \frac{d^2 U}{d z^2} = \frac{d}{d z} (\overline{u^\prime w^\prime}),
\end{align}
which can be solved via numerical integration using the no-slip BCs at the plates.\\

In addition, in the supplementary material we provide a Python code snippet, which gives the solution for the various quantities $\hat{\theta}, \hat{u},\hat{w},\langle u^\prime w^\prime \rangle$. Note that $z$ runs from $-1/2$ to $1/2$ and there is a singularity for $Pr$ $=$ $1$, which can be avoided by choosing a value very close to one or could be resolved by L'H{\^o}spital's rule.

\section{Heat and momentum transport}
 
\begin{figure}
\begin{minipage}[c]{0.995\textwidth}
\centering
\includegraphics[width=0.90\textwidth,trim=0 0 0 0,clip]{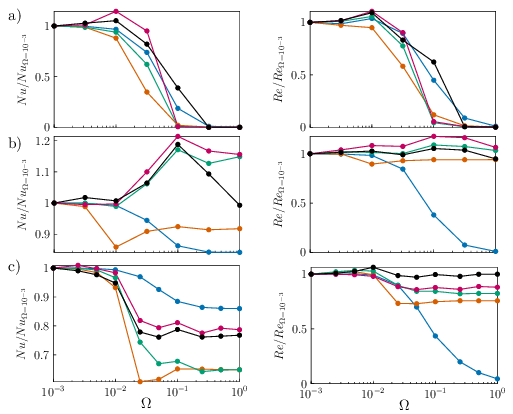}
\end{minipage}
\caption{Normalized $Nu$ and $Re$ vs. $\Omega$ for a) 2D Setup A, b) 2D Setup B and c) 3D Cylinder. $Ra =$ $10^3$ (\tikzcircle[white,fill=c1]{3pt}), $10^4$ (\tikzcircle[white,fill=c2]{3pt}), $10^5$ (\tikzcircle[white,fill=c3]{3pt}), $10^6$ (\tikzcircle[white,fill=c4]{3pt}) and $10^7$ (\tikzcircle[white,fill=c6]{3pt}).}
\label{fig:nu_re}
\end{figure}

The Nusselt number $Nu$ and Reynolds number $Re$, based on the wind velocity, are defined as
\begin{align*}
Nu \equiv - \left\langle\frac{\partial \overline{\theta}}{\partial z}\biggr\rvert_{z=0} \right\rangle_{\mathcal{A}}, \quad Re \equiv  \sqrt{\frac{Ra}{ Pr}} \sqrt{\langle \overline{\mathbf{u}^2}\rangle_{V}},
\end{align*}
where $\mathcal{A}$ denotes the horizontal plane for the cylinder or, respectively, the $x$-direction for the 2D simulations. Figure \ref{fig:nu_re} shows $Nu(\Omega)$ and $Re(\Omega)$, normalized by their values at $\Omega=10^{-3}$. Their exact values are given in the supplementary material. The 2D system (figure \ref{fig:nu_re} a,b) shows a significant heat and momentum transport enhancement for certain travelling wave speeds $\Omega$, especially for large $Ra$. For the 3D cylindrical system (figure \ref{fig:nu_re} c), no clear correlation between the zonal flow maximum (see figure \ref{fig:rbc_UvsOm}) and $Nu(\Omega)$ and $Re(\Omega)$ is observed. However a small $Re$ enhancement is present at  $\Omega\approx10^{-2}$. 

\section{Linear stability analysis}
\label{app:lst}
In section \ref{sec:2d-prograde} a temporal linear stability analysis was conducted to identify the most unstable eigenmode of the 2D linearized Navier-Stokes equations, where a wave-like form was considered only in time. Thus, any flow quantity $\phi(x,z,t)$ is represented as $\phi(x,z,t) = \hat{\phi}(x,z)e^{-i\omega t}$ and the system of equations for the horizontal velocity $u$, the vertical velocity $w$, the pressure $p$ and the temperature $\theta$ reads

\begin{equation}
\begin{bmatrix}
L_{2D} + D_x U        & D_z U          & D_x      &  0\\ 
D_x W                 & L_{2D} + D_z W & D_z      &  -1\\
D_x                   & D_z            & 0        &  0\\
D_x \overline{\theta}            & D_z \overline{\theta}     & 0       &  K_{2D}\\
\end{bmatrix}
\begin{bmatrix}
\hat{u} \\
\hat{v} \\
\hat{p} \\
\hat{\theta} 
\end{bmatrix}
=
\omega
\begin{bmatrix}
i & 0 & 0 & 0 \\ 
0 & i & 0 & 0 \\
0 & 0 & 0 & 0 \\
0 & 0 & 0 & i \\
\end{bmatrix}
\begin{bmatrix}
\hat{u} \\
\hat{v} \\
\hat{p} \\
\hat{\theta}
\end{bmatrix},
\label{eq:B_LS}
\end{equation}

where

\begin{align*}
L_{2D} &=  U D_x + W D_z + \sqrt{Pr/Ra}\left(-D_x^2-D_z^2\right) ,\\
K_{2D} &=  U D_x + W D_z + 1/\sqrt{RaPr}\left(-D_x^2-D_z^2\right).
\end{align*}

The overline represents the mean field quantity. In our study we applied Chebyshev method to approximate the vertical gradient ($D_z$) and Fourier method for the horizontal gradient ($D_x$). Conveniently, the corresponding differentiation matrices are available open source, e.g.  we used the Python package \textit{dmsuite}.

The linear set of equations (\ref{eq:B_LS}) is solved as a generalized eigenvalue problem of the form $\mathbf{A} \hat{\phi}=\omega \mathbf{B} \hat{\phi}$,  where the eigenvectors $\phi(x,z,t)$ represent the wave amplitudes and the eigenvalues $\omega$ their respective temporal behaviour. The matrices of the size $4 \times N_x \times N_z$ are very large and therefore an iterative solver has to be used (e.g. Python's \textit{scipy.eigs}). The code has been validated by solving the Blasius boundary layer, pipe flow and Rayleigh-Taylor instability in 1D and 2D, and in closed and periodic domains. For all cases we have found excellent agreement with literature results.

\end{appendix}
 

\bibliographystyle{jfm}  
\bibliography{References}  

\begin{thebibliography}{44}
\expandafter\ifx\csname natexlab\endcsname\relax\def\natexlab#1{#1}\fi
\def\au#1{#1} \def\ed#1{#1} \def\yr#1{#1}\def\at#1{#1}\def\jt#1{\textit{#1}}
  \def\bt#1{#1}\def\bvol#1{\textbf{#1}} \def\vol#1{#1} \def\pg#1{#1}
  \def\publ#1{#1}\def\arxiv#1{#1}\def\org#1{#1}\def\st#1{\textit{#1}}

\bibitem[Ahlers {\em et~al.\/}(2009)Ahlers, Grossmann \& Lohse]{Ahlers2009}
{\sc \au{Ahlers, G.}, \au{Grossmann, S.} \& \au{Lohse, D.}} \yr{2009}
  \at{{Heat transfer and large scale dynamics in turbulent Rayleigh--B\'enard
  convection}}.  \jt{Rev. Mod. Phys.}  \bvol{81},  \pg{503--537}.

\bibitem[Bensimon {\em et~al.\/}(1990)Bensimon, Croquette, Kolodner, Williams
  \& Surko]{Bensimon1990}
{\sc \au{Bensimon, D.}, \au{Croquette, V.}, \au{Kolodner, P.}, \au{Williams,
  H.} \& \au{Surko, C.~M.}} \yr{1990}  \at{{Competing and coexisting dynamical
  states of travelling-wave convection in an annulus}}.  \jt{J. Fluid Mech.}
  \bvol{217},  \pg{441--467}.

\bibitem[Bodenschatz {\em et~al.\/}(1991)Bodenschatz, de~Bruyn, Ahlers \&
  Cannell]{Bodenschatz1991}
{\sc \au{Bodenschatz, E.}, \au{de~Bruyn, J.~R.}, \au{Ahlers, G.} \&
  \au{Cannell, D.~S.}} \yr{1991}  \at{{Transitions between patterns in thermal
  convection}}.  \jt{Phys. Rev. Lett.}  \bvol{67}~(22),  \pg{3078--3081}.

\bibitem[Bodenschatz {\em et~al.\/}(2000)Bodenschatz, Pesch \&
  Ahlers]{Bodenschatz2000}
{\sc \au{Bodenschatz, E.}, \au{Pesch, W.} \& \au{Ahlers, G.}} \yr{2000}
  \at{{Recent developments in Rayleigh--B\'enard convection}}.  \jt{Annu. Rev.
  Fluid Mech.}  \bvol{32},  \pg{709--778}.

\bibitem[Busse(1972)]{Busse1972}
{\sc \au{Busse, F.~H.}} \yr{1972}  \at{{On the mean flow induced by a thermal
  wave}}.  \jt{J. Atmos. Sci.}  \bvol{29}~(8),  \pg{1423--1429}.

\bibitem[Busse(1983)]{Busse1983}
{\sc \au{Busse, F.~H.}} \yr{1983}  \at{{Generation of mean flows by thermal
  convection}}.  \jt{Phys. D Nonlinear Phenom.}  \bvol{9}~(3),  \pg{287--299}.

\bibitem[Chawla \& Purushothaman(1983)]{Chawla1983}
{\sc \au{Chawla, S.~S.} \& \au{Purushothaman, R.}} \yr{1983}  \at{Fluid motion
  induced by travelling thermal waves in a rotating fluid}.  \jt{Geophys.
  Astrophys. Fluid Dyn.}  \bvol{26},  \pg{303--320}.

\bibitem[Cross \& Tu(1995)]{Cross1995}
{\sc \au{Cross, M.~C.} \& \au{Tu, Y.}} \yr{1995}  \at{{Defect dynamics for
  spiral chaos in Rayleigh-B\'enard convection}}.  \jt{Phys. Rev. Lett.}
  \bvol{75},  \pg{834--837}.

\bibitem[Davey(1967)]{Davey1967}
{\sc \au{Davey, A.}} \yr{1967}  \at{{The motion of a fluid due to a moving
  source of heat at the boundary}}.  \jt{J. Fluid Mech.}  \bvol{29}~(1),
  \pg{137--150}.

\bibitem[Decker {\em et~al.\/}(1994)Decker, Pesch \& Weber]{Decker1994}
{\sc \au{Decker, W.}, \au{Pesch, W.} \& \au{Weber, A.}} \yr{1994}  \at{{Spiral
  defect chaos in Rayleigh-B\'enard convection}}.  \jt{Phys. Rev. Lett.}
  \bvol{73},  \pg{648--651}.

\bibitem[Favier \& Knobloch(2020)]{Favier2020}
{\sc \au{Favier, B.} \& \au{Knobloch, E.}} \yr{2020}  \at{{Robust wall states
  in rapidly rotating Rayleigh--B\'enard convection}}.  \jt{J. Fluid Mech.}
  \bvol{895},  \pg{R1}.

\bibitem[Fultz {\em et~al.\/}(1959)Fultz, Long, Owens, Bohan, Kaylor \&
  Weil]{Fultz1959}
{\sc \au{Fultz, D.}, \au{Long, R.~R.}, \au{Owens, G.~V.}, \au{Bohan, W.},
  \au{Kaylor, R.} \& \au{Weil, J.}} \yr{1959}  \at{{Studies of thermal
  convection in a rotating cylinder with some implications for large-scale
  atmospheric motions}}  \bvol{4},  \pg{1--104}.

\bibitem[Goluskin {\em et~al.\/}(2014)Goluskin, Johnston, Flierl \&
  Spiegel]{Goluskin2014}
{\sc \au{Goluskin, David}, \au{Johnston, Hans}, \au{Flierl, Glenn~R.} \&
  \au{Spiegel, Edward~A.}} \yr{2014}  \at{{Convectively driven shear and
  decreased heat flux}}.  \jt{J. Fluid Mech.}  \bvol{759}~(6),  \pg{360--385}.

\bibitem[Halley(1687)]{Halley1687}
{\sc \au{Halley, E.}} \yr{1687}  \at{{An historical account of the trade winds,
  and monsoons, observable in the seas between and near the Tropicks, with an
  attempt to assign the physical cause of the said winds}}.  \jt{Philos. Trans.
  R. Soc. London}  \bvol{16},  \pg{153--168}.

\bibitem[Hinch \& Schubert(1971)]{Hinch1971}
{\sc \au{Hinch, E.~J.} \& \au{Schubert, G.}} \yr{1971}  \at{{Strong streaming
  induced by a moving thermal wave}}.  \jt{J. Fluid Mech.}  \bvol{47}~(2),
  \pg{291--304}.

\bibitem[Howard \& Krishnamurti(1986)]{Howard1984}
{\sc \au{Howard, L.~N.} \& \au{Krishnamurti, R.}} \yr{1986}  \at{{Large-scale
  flow in turbulent convection: a mathematical model}}.  \jt{J. Fluid Mech.}
  \bvol{170},  \pg{385–410}.

\bibitem[Kelly \& Vreeman(1970)]{Kelly1970}
{\sc \au{Kelly, R.~E.} \& \au{Vreeman, J.~D.}} \yr{1970}  \at{{Excitation of
  waves and mean currents in a stratified fluid due to a moving heat source}}.
  \jt{Zeitschrift f{\"{u}}r Angew. Math. und Phys.}  \bvol{21}~(1),
  \pg{1--16}.

\bibitem[Knobloch \& Silber(1990)]{Knobloch1990}
{\sc \au{Knobloch, E.} \& \au{Silber, M.}} \yr{1990}  \at{{Travelling wave
  convection in a rotating layer}}.  \jt{Geophys. Astrophys. Fluid Dyn.}
  \bvol{51}~(1-4),  \pg{195--209}.

\bibitem[Kolodner {\em et~al.\/}(1988)Kolodner, Bensimon \&
  Surko]{Kolodner1988}
{\sc \au{Kolodner, P.}, \au{Bensimon, D} \& \au{Surko, C.~M.}} \yr{1988}
  \at{{Travelling-wave convection in an annulus}}.  \jt{Phys. Rev. Lett.}
  \bvol{60},  \pg{1723--1726}.

\bibitem[Kolodner \& Surko(1988)]{Kolodner1988b}
{\sc \au{Kolodner, P.} \& \au{Surko, C.~M.}} \yr{1988}  \at{{Weakly nonlinear
  traveling-wave convection}}.  \jt{Phys. Rev. Lett.}  \bvol{61}~(7),
  \pg{842--845}.

\bibitem[Kooij {\em et~al.\/}(2018)Kooij, Botchev, Frederix, Geurts, Horn,
  Lohse, van~der Poel, Shishkina, Stevens \& Verzicco]{Kooij2018}
{\sc \au{Kooij, G.~L.}, \au{Botchev, M.~A.}, \au{Frederix, E.~M.A.},
  \au{Geurts, B.~J.}, \au{Horn, S.}, \au{Lohse, D.}, \au{van~der Poel, E.~P.},
  \au{Shishkina, O.}, \au{Stevens, R. J. A.~M.} \& \au{Verzicco, R.}} \yr{2018}
   \at{{Comparison of computational codes for direct numerical simulations of
  turbulent Rayleigh--B\'enard convection}}.  \jt{Comp. Fluids}  \bvol{166},
  \pg{1--8}.

\bibitem[Krishnamurti \& Howard(1981)]{Krishnamurti1981}
{\sc \au{Krishnamurti, R.} \& \au{Howard, L.~N.}} \yr{1981}  \at{{Large-scale
  flow generation in turbulent convection}}.  \jt{Proc. Natl. Acad. Sci.}
  \bvol{78}~(4),  \pg{1981--1985}.

\bibitem[Malkus(1970)]{Malkus1970}
{\sc \au{Malkus, W. V.~R.}} \yr{1970}  \at{{Hadley-Halley circulation on
  Venus}}.  \jt{J. Atmos. Sci.}  \bvol{27}~(4),  \pg{529--535}.

\bibitem[Mamou {\em et~al.\/}(1996)Mamou, Robillard, Bilgen \&
  Vasseur]{Mamou1996}
{\sc \au{Mamou, M}, \au{Robillard, L}, \au{Bilgen, E} \& \au{Vasseur, P}}
  \yr{1996}  \at{{Effects of a moving thermal wave on B{\'{e}}nard convection
  in a horizontal saturated porous layer}}.  \jt{Int. J. Heat Mass Transf.}
  \bvol{39}~(2),  \pg{347--354}.

\bibitem[Maximenko {\em et~al.\/}(2005)Maximenko, Bang \&
  Sasaki]{Maximenko2005}
{\sc \au{Maximenko, N.~A.}, \au{Bang, B.} \& \au{Sasaki, H.}} \yr{2005}
  \at{Observational evidence of alternating zonal jets in the world ocean}.
  \jt{Geophys. Res. Lett.}  \bvol{32}~(12).

\bibitem[Nadiga(2006)]{Nadiga2006}
{\sc \au{Nadiga, B.~T.}} \yr{2006}  \at{On zonal jets in oceans}.  \jt{Geophys.
  Res. Lett.}  \bvol{33}~(10).

\bibitem[Niemela \& Sreenivasan(2008)]{Niemela2008}
{\sc \au{Niemela, J.~J.} \& \au{Sreenivasan, K.~R.}} \yr{2008}  \at{{Formation
  of the "superconducting" core in turbulent thermal convection}}.  \jt{Phys.
  Rev. Lett.}  \bvol{100},  \pg{184502}.

\bibitem[Schubert \& Whitehead(1969)]{Schubert1969}
{\sc \au{Schubert, G.} \& \au{Whitehead, J.~A.}} \yr{1969}  \at{{Moving flame
  experiment with liquid Mercury: Possible implications for the Venus
  atmosphere}}.  \jt{Science}  \bvol{163}~(3862),  \pg{71--72}.

\bibitem[Shishkina {\em et~al.\/}(2015)Shishkina, Horn, Wagner \&
  Ching]{Shishkina2015}
{\sc \au{Shishkina, O.}, \au{Horn, S.}, \au{Wagner, S.} \& \au{Ching, E.
  S.~C.}} \yr{2015}  \at{{Thermal boundary layer equation for turbulent
  Rayleigh--B\'enard convection}}.  \jt{Phys. Rev. Lett.}  \bvol{114},
  \pg{114302}.

\bibitem[Shishkina {\em et~al.\/}(2010)Shishkina, Stevens, Grossmann \&
  Lohse]{Shishkina2010}
{\sc \au{Shishkina, O.}, \au{Stevens, R. J. A.~M.}, \au{Grossmann, S.} \&
  \au{Lohse, D.}} \yr{2010}  \at{Boundary layer structure in turbulent thermal
  convection and its consequences for the required numerical resolution}.
  \jt{New J. Phys.}  \bvol{12},  \pg{075022}.

\bibitem[Shukla {\em et~al.\/}(1981)Shukla, Yu, Rahman \&
  Spatschek]{Shukla1981}
{\sc \au{Shukla, P.~K.}, \au{Yu, M.~Y.}, \au{Rahman, H.~U.} \& \au{Spatschek,
  K.~H.}} \yr{1981}  \at{{Excitation of convective cells by drift waves}}.
  \jt{Phys. Rev. A}  \bvol{23}~(1),  \pg{321--324}.

\bibitem[Stern(1959)]{Stern1959}
{\sc \au{Stern, M.~E.}} \yr{1959}  \at{{The moving flame experiment}}.
  \jt{Tellus}  \bvol{11},  \pg{175--179}.

\bibitem[Stern(1971)]{Stern1971}
{\sc \au{Stern, M.~E.}} \yr{1971}  \at{{Generalizations of the rotating flame
  effect}}.  \jt{Tellus}  \bvol{23},  \pg{122--128}.

\bibitem[Thompson(1970)]{Thompson1970}
{\sc \au{Thompson, R.}} \yr{1970}  \at{{Venus general circulation is a
  merry-go-round}}.  \jt{J. Atmos. Sci.}  \bvol{27}~(8),  \pg{1107--1116}.

\bibitem[Venezian(1969)]{Venezian1969}
{\sc \au{Venezian, G.}} \yr{1969}  \at{{Effect of modulation on the onset of
  thermal convection}}.  \jt{J. Fluid Mech.}  \bvol{35},  \pg{243–254}.

\bibitem[Wang {\em et~al.\/}(2012)Wang, Ma, Chen \& Sun]{Wang2012}
{\sc \au{Wang, B.~F.}, \au{Ma, D.~J.}, \au{Chen, C.} \& \au{Sun, D.~J.}}
  \yr{2012}  \at{{Linear stability analysis of cylindrical
  Rayleigh--B{\'{e}}nard convection}}.  \jt{J. Fluid Mech.}  \bvol{711},
  \pg{27--39}.

\bibitem[Wang {\em et~al.\/}(2020{\natexlab{{\em a\/}}})Wang, Chong, Stevens \&
  Lohse]{Wang2020a}
{\sc \au{Wang, Q.}, \au{Chong, K.~L.}, \au{Stevens, R.} \& \au{Lohse, D.}}
  \yr{2020{\natexlab{{\em a\/}}}}  \at{{From zonal flow to convection rolls in
  Rayleigh--B\'enard convection with free-slip plates}}.  \jt{J. Fluid Mech.}
  \bvol{905}.

\bibitem[Wang {\em et~al.\/}(2020{\natexlab{{\em b\/}}})Wang, Verzicco, Lohse
  \& Shishkina]{Wang2020}
{\sc \au{Wang, Q.}, \au{Verzicco, R.}, \au{Lohse, D.} \& \au{Shishkina, O.}}
  \yr{2020{\natexlab{{\em b\/}}}}  \at{Multiple states in turbulent
  large-aspect-ratio thermal convection: What determines the number of
  convection rolls?}  \jt{Phys. Rev. Lett.}  \bvol{125},  \pg{074501}.

\bibitem[Whitehead(1972)]{Whitehead1972}
{\sc \au{Whitehead, J.~A.}} \yr{1972}  \at{{Observations of rapid mean flow
  produced in mercury by a moving heater}}.  \jt{Geophys. Fluid Dyn.}
  \bvol{3}~(2),  \pg{161--180}.

\bibitem[Whitehead(1975)]{Whitehead1975}
{\sc \au{Whitehead, J.~A.}} \yr{1975}  \at{{Mean flow generated by circulation
  on a beta-plane: An analogy with the moving flame experiment}}.  \jt{Tellus}
  \bvol{27}~(4),  \pg{358--364}.

\bibitem[Yang {\em et~al.\/}(2020)Yang, Chong, Wang, Verzicco, Shishkina \&
  Lohse]{Yang2020}
{\sc \au{Yang, R.}, \au{Chong, K.~L.}, \au{Wang, Q.}, \au{Verzicco, R.},
  \au{Shishkina, O.} \& \au{Lohse, D.}} \yr{2020}  \at{{Periodically modulated
  thermal convection}}.  \jt{Phys. Rev. Lett.}  \bvol{125},  \pg{154502}.

\bibitem[Yano {\em et~al.\/}(2003)Yano, Talagrand \& Drossart]{Yano2003}
{\sc \au{Yano, J.}, \au{Talagrand, O.} \& \au{Drossart, P.}} \yr{2003}
  \at{{Origins of atmospheric zonal winds}}.  \jt{Nature}  \bvol{36}.

\bibitem[Young {\em et~al.\/}(1972)Young, Schubert \& Torrance]{Young1972}
{\sc \au{Young, R.~E.}, \au{Schubert, G.} \& \au{Torrance, K.~E.}} \yr{1972}
  \at{{Nonlinear motions induced by moving thermal waves}}.  \jt{J. Fluid
  Mech.}  \bvol{54}~(1),  \pg{163--187}.

\bibitem[Zhang {\em et~al.\/}(2020)Zhang, van Gils, Horn, Wedi, Zwirner,
  Ahlers, Ecke, Weiss, Bodenschatz \& Shishkina]{Zhang2020}
{\sc \au{Zhang, X.}, \au{van Gils, D. P.~M.}, \au{Horn, S.}, \au{Wedi, M.},
  \au{Zwirner, L.}, \au{Ahlers, G.}, \au{Ecke, R.~E.}, \au{Weiss, S.},
  \au{Bodenschatz, E.} \& \au{Shishkina, O.}} \yr{2020}  \at{{Boundary zonal
  flow in rotating turbulent Rayleigh--B\'enard convection}}.  \jt{Phys. Rev.
  Lett.}  \bvol{124},  \pg{084505}.

\end{thebibliography}

\end{document}